\documentclass{aa}  

\newcommand{\twcooz}{$^{12}$CO $(J=1-0)$}
\newcommand{\twcoto}{$^{12}$CO $(J=2-1)$}
\newcommand{\thcooz}{$^{13}$CO $(J=1-0)$}

\usepackage{xcolor}
\newcommand \msun{\(M_\odot\)}

\usepackage{graphicx}
\usepackage{txfonts}
\begin{document}

   \title{Two shell- and wing-shaped supernova remnants.}

   \subtitle{Investigating the molecular environments around VRO 42.05.01 and G 350.0-2.0}

   \author{Maria Arias
          \inst{1}
          \and
          Ping Zhou
          \inst{2}
          \and
          Alexandros Chiotellis
          \inst{3}
            \and
          Carlos De Breuck
          \inst{4}
          \and
          Vladimir Dom\v{c}ek
          \inst{5}
          \and
          Panos Boumis
          \inst{3}
          \and
          Jacco Vink
          \inst{5}
          \and
          Sophia Derlopa
          \inst{3}
          \and
          Stavros Akras
          \inst{3}
          }

   \institute{Leiden Observatory, Leiden University, PO Box 9513, 2300 RA, Leiden, The Netherlands \\
              \email{arias@strw.leidenuniv.nl}
              \and
              School of Astronomy and Space Science, Nanjing University, Nanjing 210023, PR China
              \and
        Institute for Astronomy, Astrophysics, Space Applications and Remote Sensing, National Observatory of Athens, GR 15236 Penteli, Greece 
        \and
    European Southern Observatory, Karl-Schwarzschild-Str. 2 85748 Garching bei München, Germany                       
    \and    
    Anton Pannekoek Institute for Astronomy \& GRAPPA, University of Amsterdam, Science Park 904, 1098 XH Amsterdam, The Netherlands
    }
   \date{}

 
  \abstract
  {Supernova remnants (SNRs) are profoundly affected by their ambient medium. In particular, SNRs with a mixed morphology (characterised by a shell-like radio morphology and centrally filled X-ray emission) are thought to be the result of the interaction of a supernova explosion with a dense environment. 
  In this work, we present carbon monoxide (CO) observations around two mixed morphology SNRs, VRO 42.05.01 and G 350.0-2.0, that look remarkably similar in continuum radio emission, showing what we refer to as a shell and wing shape. It has been proposed that the shell and wing shape is the result of environmental effects, in the form of a sharp density gradient or discontinuity. Therefore, our motivation for studying these two sources jointly is that if the dense molecular environment causes the development of these sources' shell and wing shape, then these two sources' environments must be similar. This is contrary to what we observe.
  In the case of VRO 42.05.01, we have found direct evidence of an interaction with its molecular environment, in the form of broadened $^{12}$CO line profiles, high \twcoto\ to \twcooz\ line ratios, and arc features in position-velocity space. We interpret some of these features to be associated with the SNR shock, and some of them to be due to the presence of a pre-supernova stellar wind.
  We have found no such features in the abundant molecular gas surrounding G 350.0-2.0. In addition to the spectral line analysis, we have used radio continuum data to make a spectral index map of G 350.0-2.0, and we see
that the radio spectrum of G 350.0-2.0 steepens significantly at frequencies $<200$~MHz, much like that of VRO 42.05.01. In spite of their spectral and morphological similarities, these two sources  look substantially different in their observed optical and infrared emission. The lack of large-scale correspondence between the radio continuum and the molecular material, in either case, as well as the differences in the excitation and morphological properties of the molecular gas surrounding both sources, lead us to conclude that the shell and wing morphology of these two sources is not due to interactions with a similar ambient molecular interstellar medium. }
  

   \maketitle
%
\section{Introduction}

Supernova remnants (SNRs) result from the interaction of an exploded star with its surrounding interstellar medium (ISM). As such, they are shaped both by the nature of the supernova explosion and by the ambient medium they encounter, which, in turn, can be perturbed by the radiation and mass-loss history of the supernova progenitor star.
Mixed morphology SNRs remnants \citep{rho98} are characterised by limb-brightened radio emission that surrounds diffuse, thermal X-ray emission. The mechanism by which SNRs develop their mixed morphology remains a matter of discussion in the field 
\citep{white91,cox99,shelton99,chen08,ohnishi11}; most models invoke a dense medium, which agrees with the fact that 15 Galactic mixed morphology SNRs \cite[out of a total of 24 catalogued,][]{vink12} are interacting with their molecular environments \citep{jiang10}.

\begin{figure*}
\centering
\includegraphics[width=0.86\columnwidth]{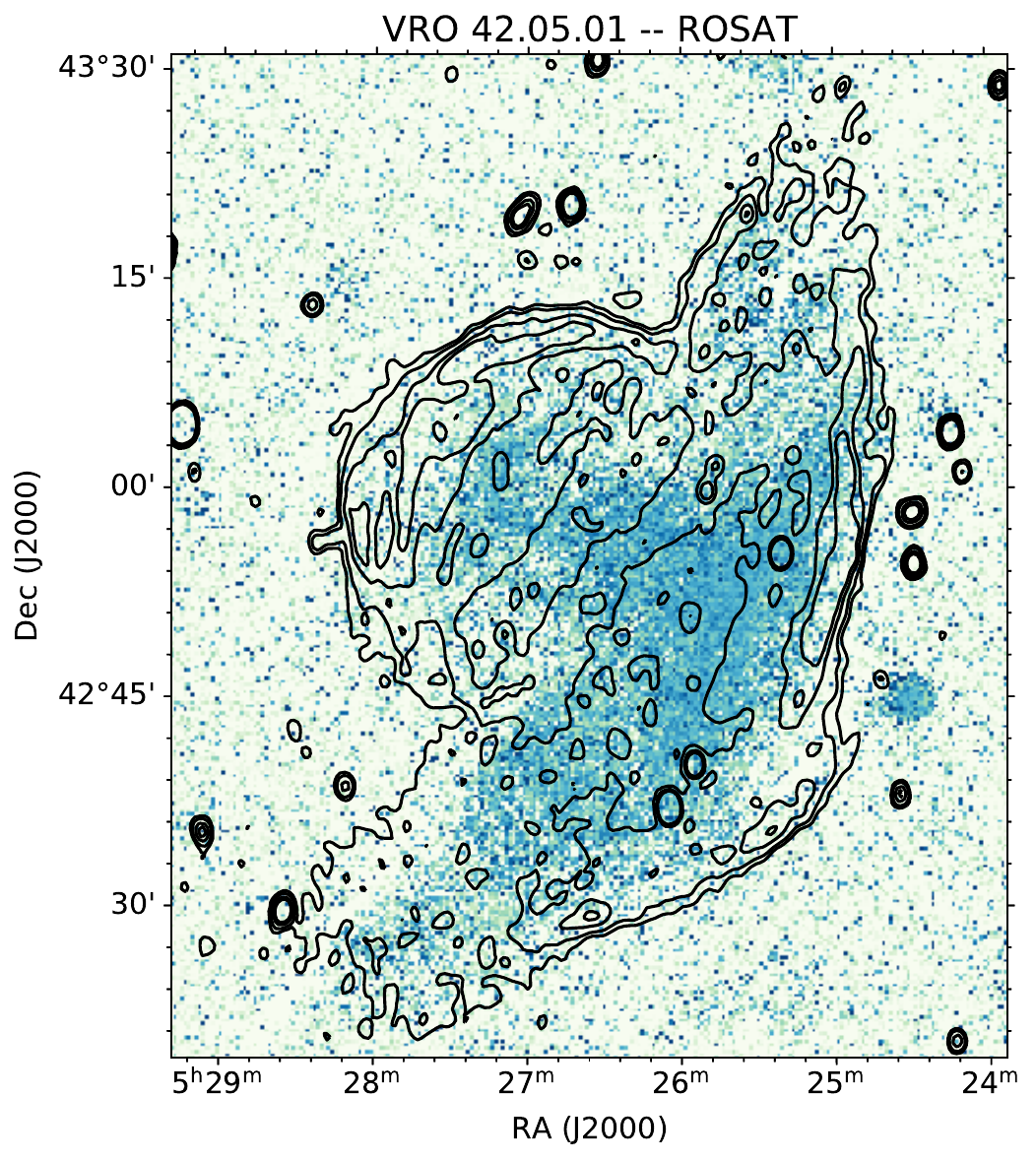}
\includegraphics[width=\columnwidth]{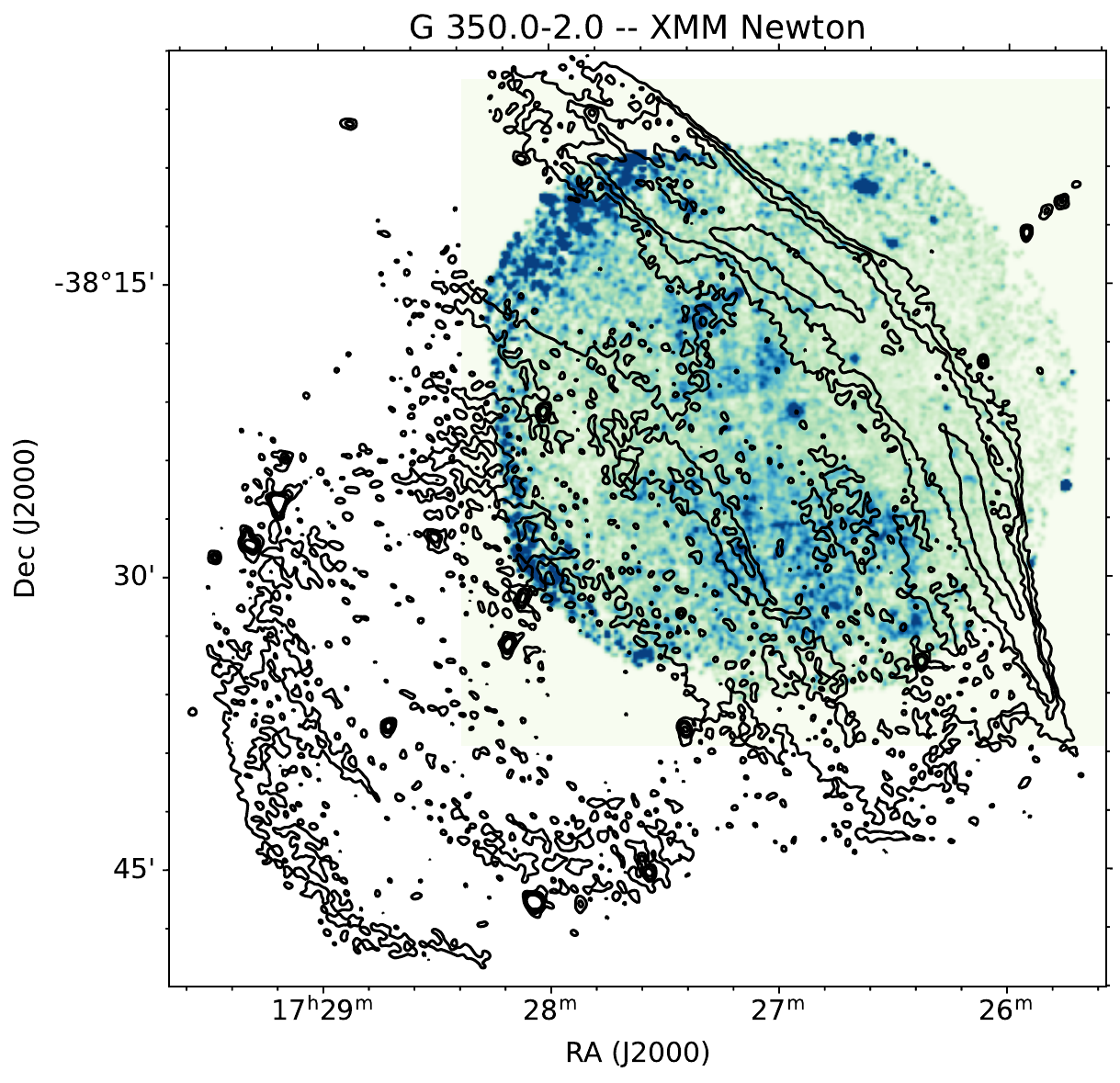}
\caption{Two shell- and wing-shaped SNRs. \textit{Left:} VRO 42.05.01 as seen with \textit{ROSAT} in the $0.1-2.4$~keV energy band \citep{voges99}. The contours are from the 1420~MHz the Canadian Galactic Plane Survey image of VRO 42.05.01 \citep{wright10}. \textit{Right:} G 350.0-2.0 as seen with the \textit{XMM-Newton} EPIC camera in the $0.5-1$~keV band. The contours are from the 1.4~GHz radio map in \cite{gaensler98}. The \textit{ROSAT} map of this source \cite[see fig. 1 in][]{karpova16} covers the entire SNR, showcasing that 350.0-2.0 is indeed a mixed morphology SNR.}
\label{fig:rgb}
\end{figure*}

VRO 42.05.01 (also known as G~166.0+4.3) and G 350.0-2.0 are two mixed morphology SNRs that look strikingly similar\footnote{The similarity is seen most clearly in the L-band radio maps --compare fig. 1 of \cite{arias19} with fig. 2 of \cite{gaensler98}.}: both have a semicircular shell, and a larger, triangular wing \cite[see Fig. \ref{fig:rgb}; we maintain the naming convention used by][originallly refererring to VRO 42.05.01]{pineault85}. For both sources the vertical line separating the shell and the wing is parallel to the Galactic plane, and the shell's semicircle is in the direction away from the Galaxy, whereas the wing points towards the Galaxy. Our motivation in jointly studying these two sources is that if the reason for developing a shell and wing morphology is environmental, then these \lq twins\rq\ should have similar environments.

VRO 42.05.01 has been studied extensively at radio, optical, and X-ray wavelengths \citep{lozinskaia79,fesen83,pineault85,burrows94,guo97,leahy05,araya13,arias19b,xiao22}. 
We proposed a distance of $1.0\pm0.4$~kpc, corresponding to 13~pc in size \citep{arias19}, and its age is approximately 7000~yr \citep{xiao22,matsumura17}.
Much of the literature on this source concerns its strange shell and wing morphology \citep{landercker82,pineault87,landecker89,seung-urn97}, which some authors \citep{pineault85,pineault87,landecker89} have suggested is due to the breakout of a part of the SNR into a low-density medium. Our earlier works \citep{arias19,arias19b,Chiotellis2019,derlopa20} proposed that pre-explosion properties of the progenitor star (rather than, or in addition to, properties of the surrounding ISM) could account for the shell and wing morphology.

In spite of its stunning similarity with VRO~42.05.01, the literature on G 350.0-2.0 is limited. It was observed at 1.4~GHz by \cite{gaensler98}, who noted that a significant number of SNRs (including G 350.0-2.0 and VRO~42.05.01) are aligned with the Galactic plane along one axis. \cite{stupar11} observed this source in H$\alpha$, where it shows no morphological similarity with the radio emission (unlike VRO 42.05.01, whose H$\alpha$ emission outlines the radio morphology). The authors identify some elongated filaments that could be associated with the SNR, and favour an interpretation whereby the source is actually more than one SNR aligned along the line of sight. \cite{karpova16} observed G 350.0-2.0 with \textit{XMM-Newton}, finding a plasma with mostly solar abundances and some regions of overabundant Fe. They estimated a distance of approximately 3~kpc, corresponding to a SNR diameter of 20~pc, and an age of $\sim10^4$~years. 

In this work, we aim to examine the role of these sources' molecular environments  in accounting for their shell and wing shape. We looked for signatures of interaction between the SNR shock and the molecular material, and for signs of the supernova progenitor from its imprint on the molecular clouds that surround it. 
We performed single-dish millimetre observations with the Atacama Pathfinder Experiment (APEX) and the IRAM 30~m telescope of G 350.0-2.0 and VRO 42.05.01, respectively, in order to map the CO emission surrounding the remnants. In Section~\ref{sec:obs}, we describe the molecular line data we took with the APEX and IRAM 30~m telescopes, as well as the archival data we have made use of for this paper. In Section~3, we present evidence that the molecular material is interacting with VRO 42.05.01's SNR shock and stellar winds from its progenitor star, which consists of ratios of \twcoto\ to \twcooz\ larger than one, broadened line velocities, and sweeping features in position-velocity space. In section 4, we present the results from the APEX observations of G 350.0-2.0, which are inconclusive in terms of confirming or rejecting interaction with the SNR shock. In section 5, we compare the two sources' infrared emission, discuss the implications of our molecular results, and note similarities and differences with other mixed morphology SNRs interacting with their neighbouring molecular gas. In section 6, we summarise our work and present our conclusions.
 
\section{Observations}\label{sec:obs}

\subsection{IRAM 30~m observations}

\begin{figure}
\centering
\includegraphics[width=0.9\columnwidth]{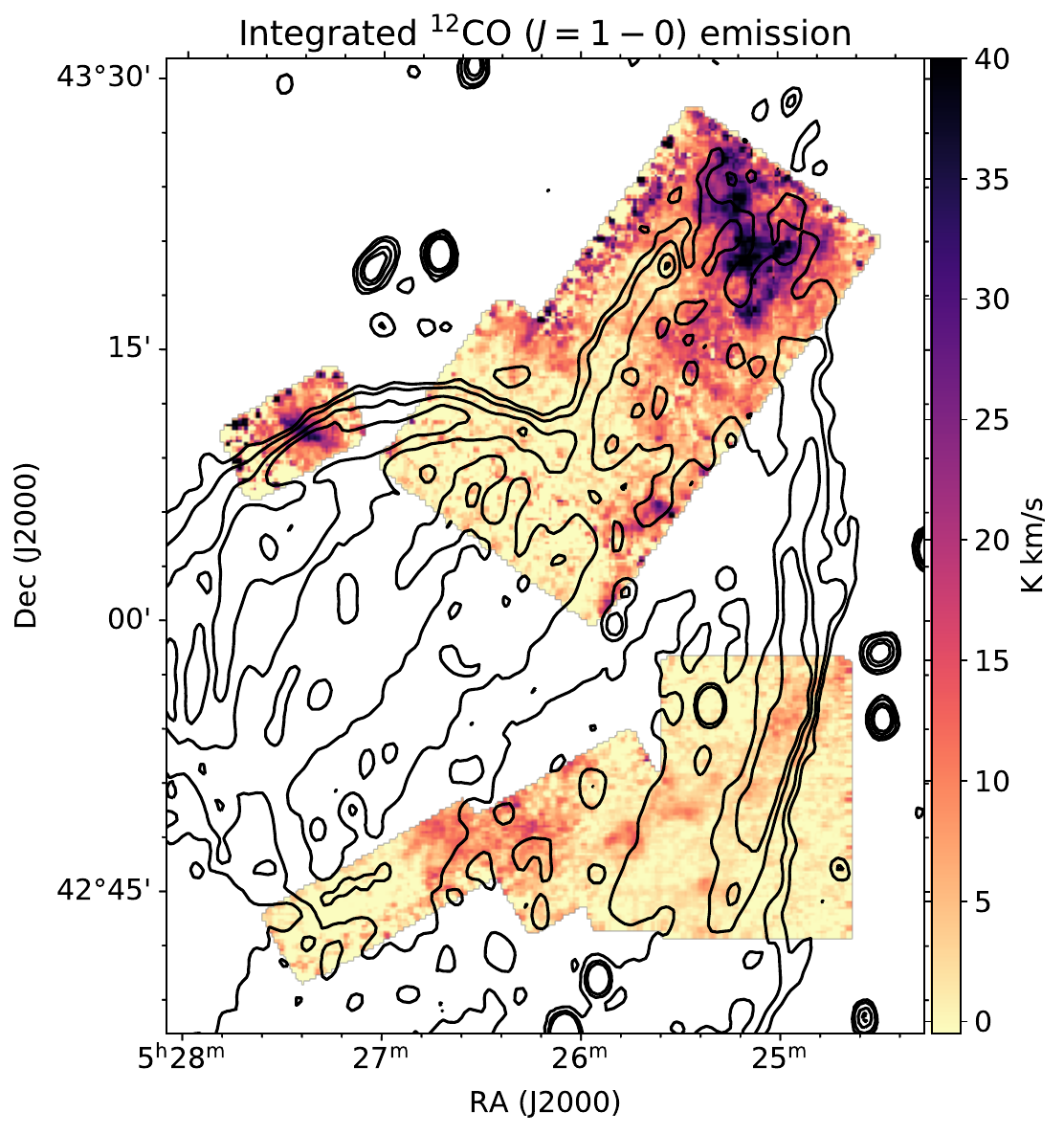}
\includegraphics[width=0.9\columnwidth]{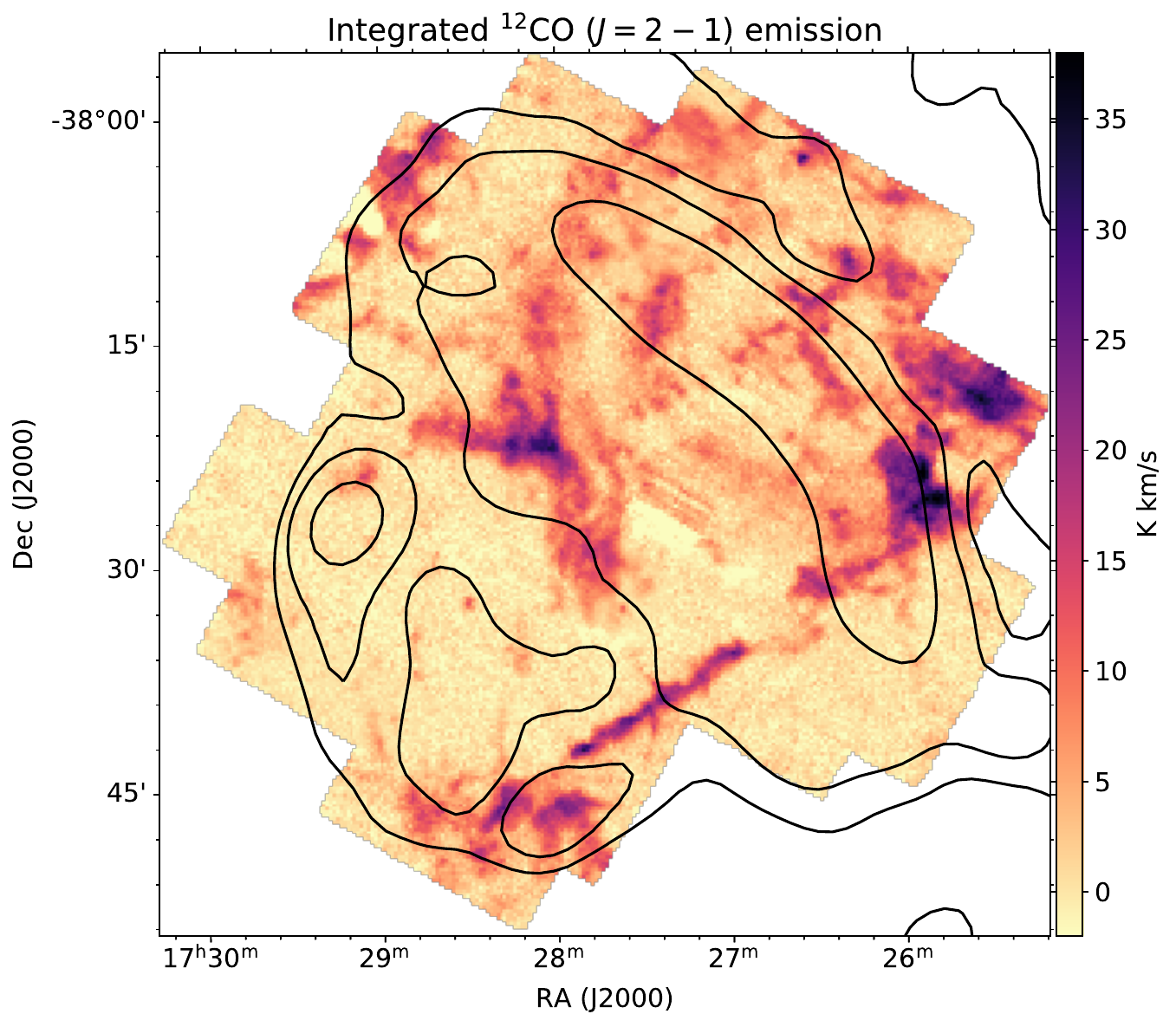}
\caption{Molecular observations towards the shell- and wing-shaped SNRs. \textit{Top:} Footprint of the IRAM observations of VRO 42.05.01, including the observations from this work (the continuous area in the southern half of this map) and those in \citet[][the two separate regions in the northern half]{arias19}. The velocity-integrated emission is shown for the \twcooz\ transition. The contours are from the 1420~MHz Canadian Galactic Plane Survey image of VRO 42.05.01 \citep{wright10}. \textit{Bottom:} Integrated spectra for the APEX G 350.0-2.0 observations, \twcoto\ transition. The radio contours are from the 88~MHz GLEAM image of G 350.0-2.0 \citep{hurley-walker19}.
}
\label{fig:footprint}
\end{figure}

We performed 3~mm and 1~mm heterodyne IRAM 30~m observations toward the south-western wing of SNR VRO 42.05.01 on March 8-10 2020 (project ID: 159-19). The observational set-up was similar to that for the observations presented in \cite{arias19}, the only difference is the area covered. In Fig.~\ref{fig:footprint} we show the footprint of the observations in this work alongside those published in \cite{arias19}.

We simultaneously mapped \twcooz\ at 115.271 GHz, \thcooz\ at 110.201 GHz, and \twcoto\ at 230.540 GHz emission towards the centre of VRO 42.05.01 in the on-the-fly (OTF) mode with the Eight MIxer Receiver (EMIR). We performed the data reduction with the GILDAS/CLASS package\footnote{\url{http://www.iram.fr/IRAMFR/GILDAS/}}, using beam efficiencies of 0.78, 0.59, and 0.78 for the \twcooz, \twcoto, and \thcooz\ transitions, respectively. 

The data that we used for analysis were in the form of \twcooz, \twcoto, and \thcooz\ main beam temperature ($T_\mathrm{mb}$) cubes with a grid spacing of 11.75\arcsec\ and a velocity resolution of 0.5~km~s$^{-1}$. The final map has 300 channels, spanning the velocity range from $-99.5$~km~s$^{-1}$ to $+50$~km~s$^{-1}$. The velocity-integrated map of the observations for the \twcooz\ transition is shown in Fig. \ref{fig:footprint}, top panel. Here the footprint of these observations (south) is shown alongside the observations from \cite{arias19}.

\subsection{APEX observations}

The heterodyne 1~mm observations of the environment around G~350.0-2.0 were performed as a Service Mode run in April, June, and October 2021 as part of ESO programme 105.209F. We observed \twcoto\ at 230.540 GHz with the Band 6 receiver in the  nFLASH instrument. The on-the-fly data were reduced according to the standard GILDAS/CLASS script used for APEX observations by the ESO operations scientists, which, for OTF maps, results in science-ready data products (ESO Operations Helpdesk, private communication). 

The resulting data product was a \twcoto\ main beam temperature cube with a grid spacing of 14\arcsec. The data have a total of 100 channels, covering local standard of rest velocities between $-50$~km~s$^{1}$ and $50$~km~s$^{-1}$ with a resolution of $1$~km~s$^{-1}$. The velocity-integrated map of the observations is shown in Fig.~\ref{fig:footprint}, bottom panel. 

\subsection{Ancillary data}

In order to show the mixed morphology nature of these sources in Fig.~\ref{fig:rgb}, we used archival X-ray and radio data. For VRO~42.05.01 we used the \textit{ROSAT} All-Sky Survey \citep{voges99} and radio data from the Canadian Galactic Plane Survey \cite[CGPS,][]{taylor03}. For G 350.0-2.0 we used data from the \textit{XMM-Newton} EPIC camera: we mosaiced the pn and MOS 1 and 2 images, and employed the standard pipeline images and exposure maps. The radio contours are from \cite{gaensler98}. The contours shown in the molecular line figures for G 350.0-2.0 (Figs.~\ref{fig:footprint}, bottom panel, and \ref{fig:G350gaussians}) are at 88~MHz and taken from the Galactic and Extragalactic All-sky Murchinson Widefield Array Survey \cite[GLEAM,][]{hurley-walker19}. In order to understand the SNRs' relation to their dust environments, we used data from the Wide-Field Infrared Survey Explorer all-sky survey \cite[WISE][]{wright10} at wavelengths of 4.6 and 12~$\mu$m. 

\begin{figure}
\includegraphics[width=\columnwidth]{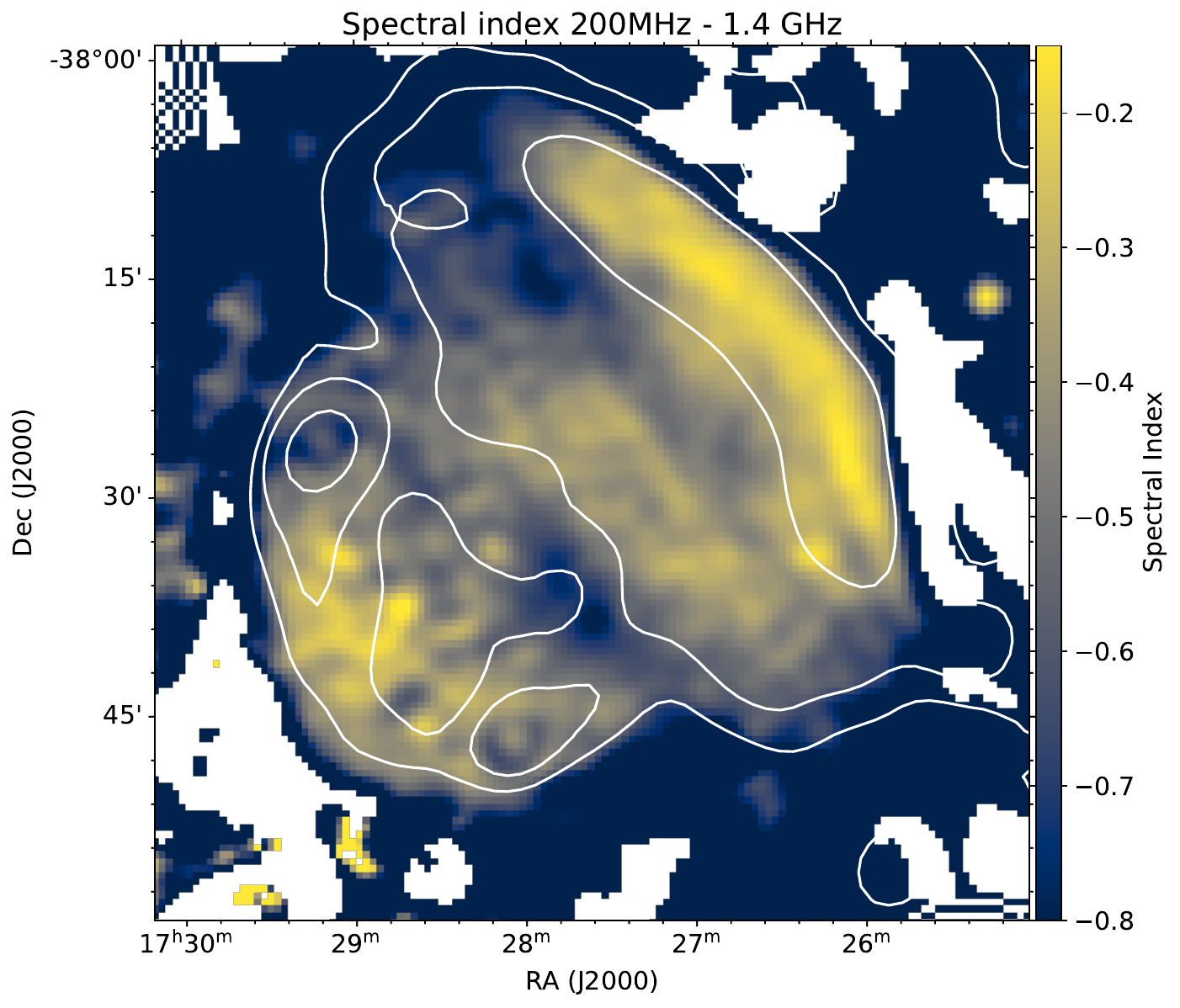}
\caption{Spectral index map of G 350.0-2.0 made from the 1.4~GHz map in \cite{gaensler98} and the GLEAM 200~MHz map \citep{hurley-walker19}. The overlaid contours are from the 88~MHz GLEAM image of G 350.0-2.0 \citep{hurley-walker19}. The visibilities are not uv-matched, nor have the point sources been subtracted, but both images are sensitive to all scales of emission for this source. The integrated spectral index value between these two frequencies is $\alpha = -0.4$. 
}
\label{fig:g30_spx}
\end{figure}

For completeness, we made a spectral index map of G 350.0-2.0, (shown in Fig. \ref{fig:g30_spx}) using the 1.4~GHz map from \cite{gaensler98} and the 200~MHz map from the GLEAM survey \citep{hurley-walker19}. The data are not uv-matched, since we did not have access to the visibilities in either case. The 1.4~GHz map contains combined data from the Very Large Array and the Parkes single-dish telescope, and the largest angular scale for the GLEAM survey is $10^\mathrm{o}$ at 227~MHz; since G 350.0-2.0 is $0.9^ \mathrm{o}$ in diameter, all angular scales are recovered in both cases and the spectral index map can be safely made. The  GLEAM map shows a gradient in the background due to the emission from the Galactic Plane (located parallel to the NE-SW axis and north of the remnant) that is not accounted for in the background subtraction, so the spectral index map does show a systematic effect due to this.

As reported by \cite{gaensler98}, G 350.0-2.0 has an integrated flux density at 1.4~GHz of $S_\mathrm{1.4~GHz} = 22.3\pm0.3$~Jy. In order to measure the flux density of G 350.0-2.0 in the GLEAM 200~MHz map, we drew a polygon around the source and subtracted the background (taking into account the gradient mentioned above). We measured an integrated flux density of $S_\mathrm{200~MHz} = 51.0\pm5.1$~Jy \cite[the 10\% error due to the intrinsic uncertainty of the low-frequency flux scale,][]{scaife12}. This results in a spectral index value between 200~MHz and 1.4~GHz of $\alpha = -0.41\pm0.06$ (see Fig.~\ref{fig:g30_spx}), coinciding with the value previously reported by \cite{gaensler98}. 

\section{Results -- VRO 42.05.01}

The bulk of emission in the CO data cubes shows at two different velocity ranges: between velocities of $-25$~km~s$^{-1}$ to $-14$~km~s$^{-1}$, and between $-6.5$~km~s$^{-1}$ to $6.5$~km~s$^{-1}$, with nearly (but not completely) empty channels in between. These different velocity components were also seen in the observations presented in \cite{arias19}, although in that case the emission occurred between $-28$~km~s$^{-1}$ to $-12$~km~s$^{-1}$, and between $-8$~km~s$^{-1}$ to $-1$~km~s$^{-1}$. In \cite{arias19} we refer to these as the \lq distant\rq\ and \lq close\rq\ components, and we maintain that naming convention here. The presence of an H {\sc i} cavity at $-6$~km~s$^{-1}$, and the existence of a velocity gradient in the close component (which we attributed to a stellar wind from VRO 42.05.01's progenitor) led us to conclude that VRO 42.05.01 is associated with the close molecular component and has a  local standard of rest (LSR) velocity of $-6$~km~s$^{-1}$. Given that the rotation curve of \cite{reid14} places the emission from the close and distant components kiloparsecs away from each other \cite[see fig. 10 in][]{arias19}, we also concluded that the emission from the distant component is unrelated to VRO 42.05.01. We will now present evidence that the data show signatures of heating in both the close and distant molecular components; however, the new data remain consistent with a LSR velocity of $-6$~km~s$^{-1}$ for the SNR.


\subsection{Evidence for interaction between the SNR shock and the molecular clouds}
\begin{figure}
\centering
\includegraphics[width=\columnwidth]{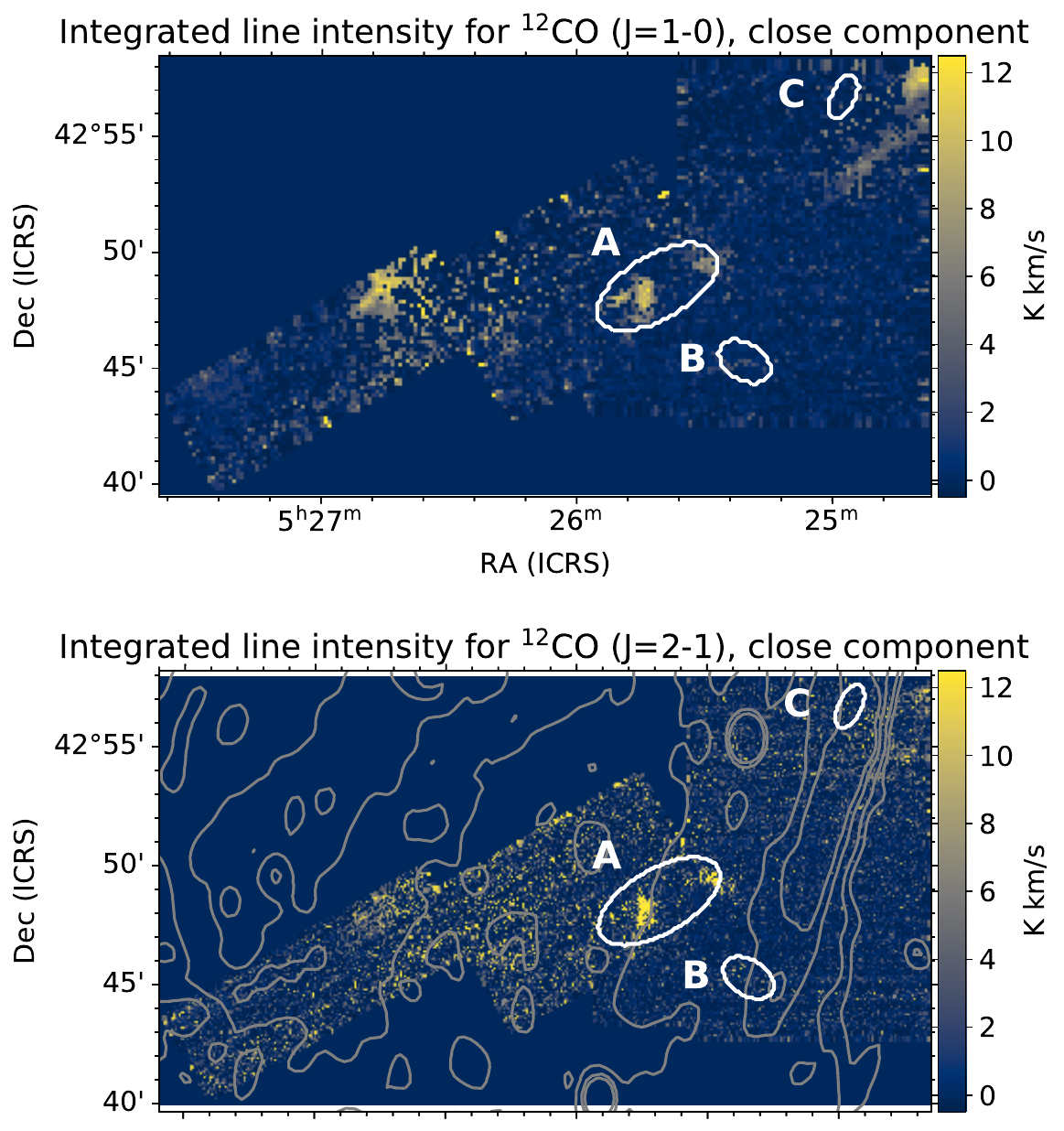}

\vspace{0.5cm}

\includegraphics[width=\columnwidth]{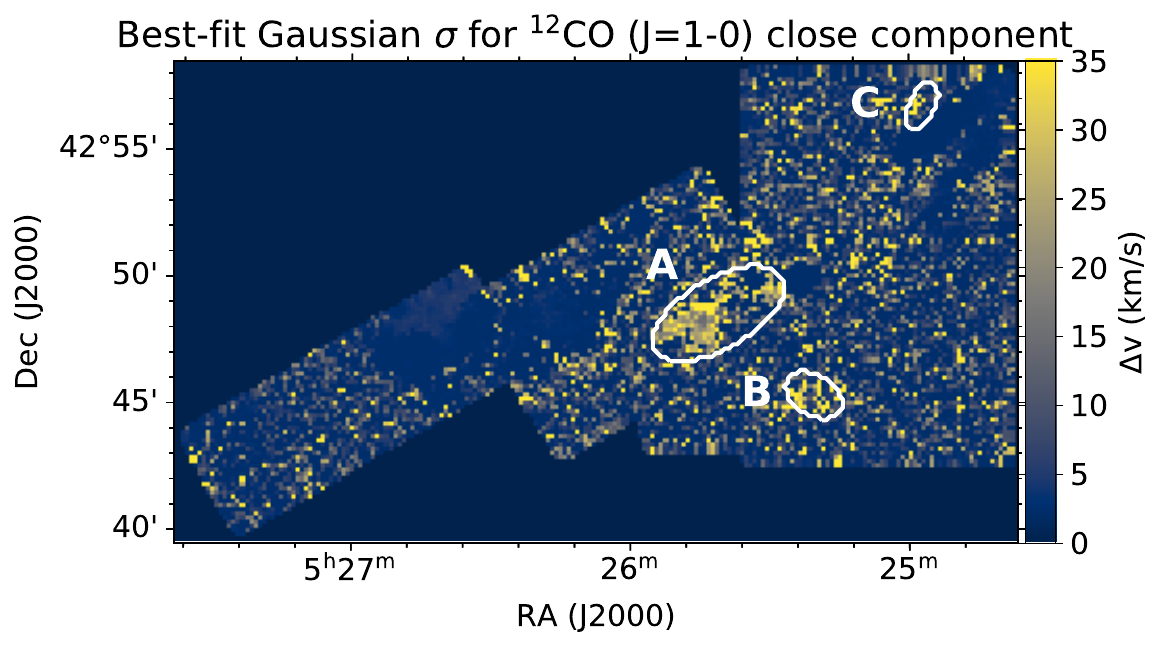}
\caption{Properties of the \twcooz\ close emission component. \textit{Top and middle:} integrated line intensities per pixel in the close component for the \twcooz\ and the \twcoto\ data, respectively, for the SNR VRO 42.05.01. The regions indicated in white mark the locations where the molecular gas is heated. \textit{Bottom:} best-fit result for the standard deviation after fitting a Gaussian to the close emission component.}
\label{fig:line_intensities}
\end{figure}

\begin{table*}[]
\centering
\begin{tabular}{cccccccc}
Region &
   RA &
   Dec &
  \begin{tabular}[c]{@{}l@{}}Central velocity\\ (km~s$^{-1}$)\end{tabular} &
  \begin{tabular}[c]{@{}l@{}}FWMH\\ (km~s$^{-1}$)\end{tabular} &
  Ratio &
  \begin{tabular}[c]{@{}l@{}}Signal-to-noise ratio \\ of \twcooz\end{tabular} &
  \begin{tabular}[c]{@{}l@{}}Signal-to-noise ratio \\ of \twcoto\end{tabular} \\ \hline

C  & 5:25:53 & +43:07:58 & -5.0  & 2.5  & 1.4 & 5.7 & 6.9  \\
Sw 3 & 5:25:57 & +43:03:39 & -7.0 & & & & \\
Sw 1 & 5:25:44 & +43:07:30 & -7.0 & & & & \\
A*/ Sw 2 & 5:25:41  & +42:48:30 & -7.5  & 13.5 & 2.0 & 9.4 & 9.7  \\
B*  & 5:25:20 & +42:45:17 & -10.0 & 11.5 & 1.8 & 7.3 & 8.9 \\
D*  & 5:24:57 & +42:56:41 & -14.0 & 10.5  & 1.9 & 4.1 & 4.8  \\
E  & 5:25:34 & +43:08:43 & -18.0 & 4.5  & 1.1 & 3.4 & 2.9  \\
G  & 5:25:34 & +43:20:59 & -21.5 & 3.0  & 1.5  & 3.0  & 4.0  \\
F  & 5:25:29 & +43:17:20 & -22.5 & 2.0  & 1.1 & 6.3 & 5.4  \\
\hline
U  & 5:25:21 & +43:22:35 & -21.5 & 1.0 & 0.65 & 119.2  & 70.2  \\
\hline
\end{tabular}
\caption{Regions towards VRO 42.05.01 that show signatures of heating, possibly due to an interaction between the SNR and the molecular cloud. Region U is shown as an example of unperturbed emission. The regions indicated with an asterisk show the clearest evidence that the SNR shock is interacting with the molecular gas. The Ratio column shows the ratio of \twcoto\ to \twcooz\ peaks in the given region. The signal-to-noise ratio columns correspond to the ratio of the peak measured for a given transition to the noise in the empty channels of the same region.}
\label{table:vro}
\end{table*}

The work of \cite{seta98} showed that an enhanced high-$J$ to low-$J$ $^{12}$CO line ratio is evidence that the emitting molecular material is heated, which can be an indication that it is interacting with the SNR shock. In order to identify regions with a high \twcoto\ to \twcooz\ ratio in VRO 42.05.01's molecular environment, we calculated the integrated line intensity, for each pixel, for of the close and distant components in the OTF maps of each of the $^{12}$CO transitions, and then took the ratio between the two maps. We found the limits of integration for the line velocities by fitting a Gaussian (within the velocity range of the either the close or distant component), and integrated numerically within $\pm 3\sigma$ of the mean using the trapezoidal rule. The results for the close component are shown in Fig.~\ref{fig:line_intensities}, top and middle. Furthermore, we plot the best-fit standard deviation for the fitted Gaussian (Fig.~\ref{fig:line_intensities}, bottom). 

\begin{figure*}
\centering
\includegraphics[width=\textwidth]{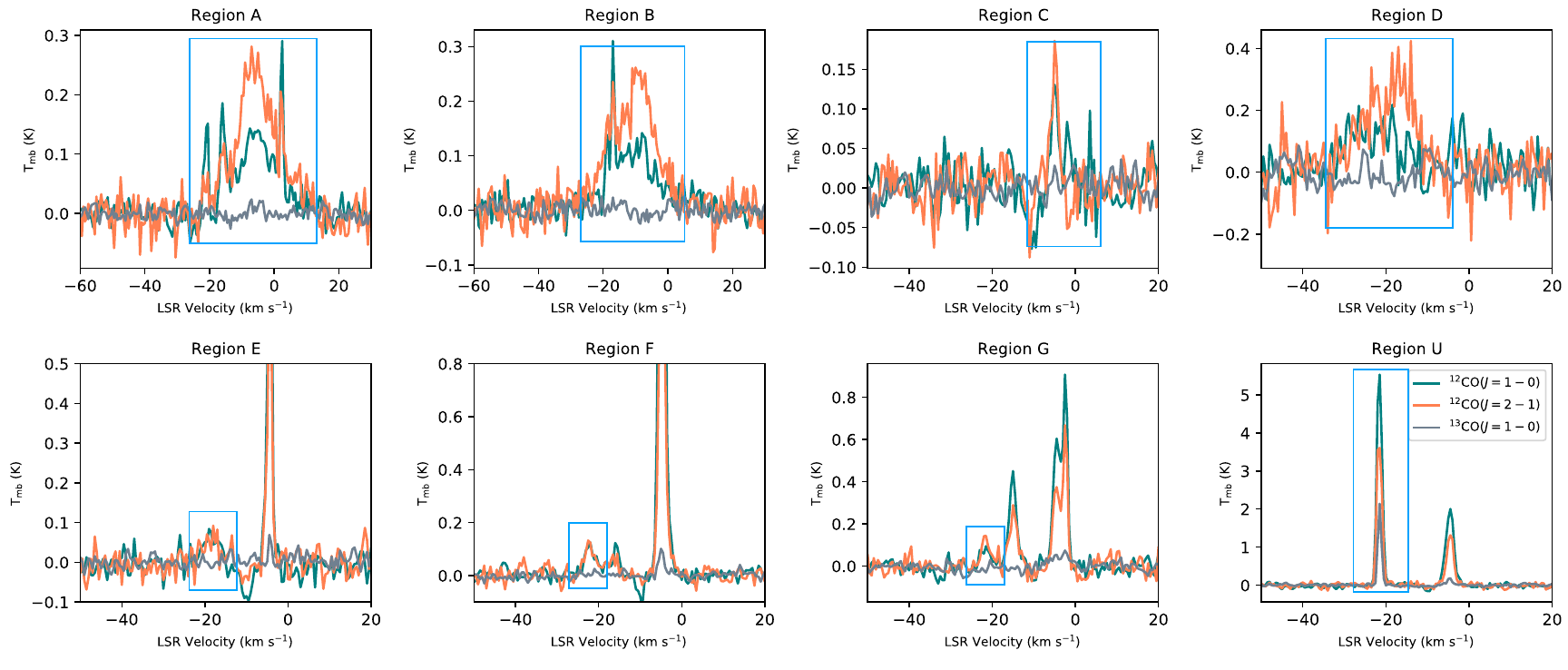}
\caption{Regions of the disturbed molecular material around VRO 42.05.01. Integrated spectra for each of the regions described in table \ref{table:vro}. The details given in the table, for each region, correspond to the line component highlighted in the blue box.
We note that regions A, B, and D show the strongest evidence for interaction with their broad velocity dispersion (FWHM) and enhanced \twcoto\ to \twcooz\ emission. Region U is shown here for comparison, as an example of undisturbed emission. The location of the regions is shown in Fig. \ref{fig:inverted_lines_locations}. The legend in the bottom, right plot is common to all plots.
}
\label{fig:inverted_lines}
\end{figure*}

\begin{figure}
\centering
\includegraphics[width=\columnwidth]{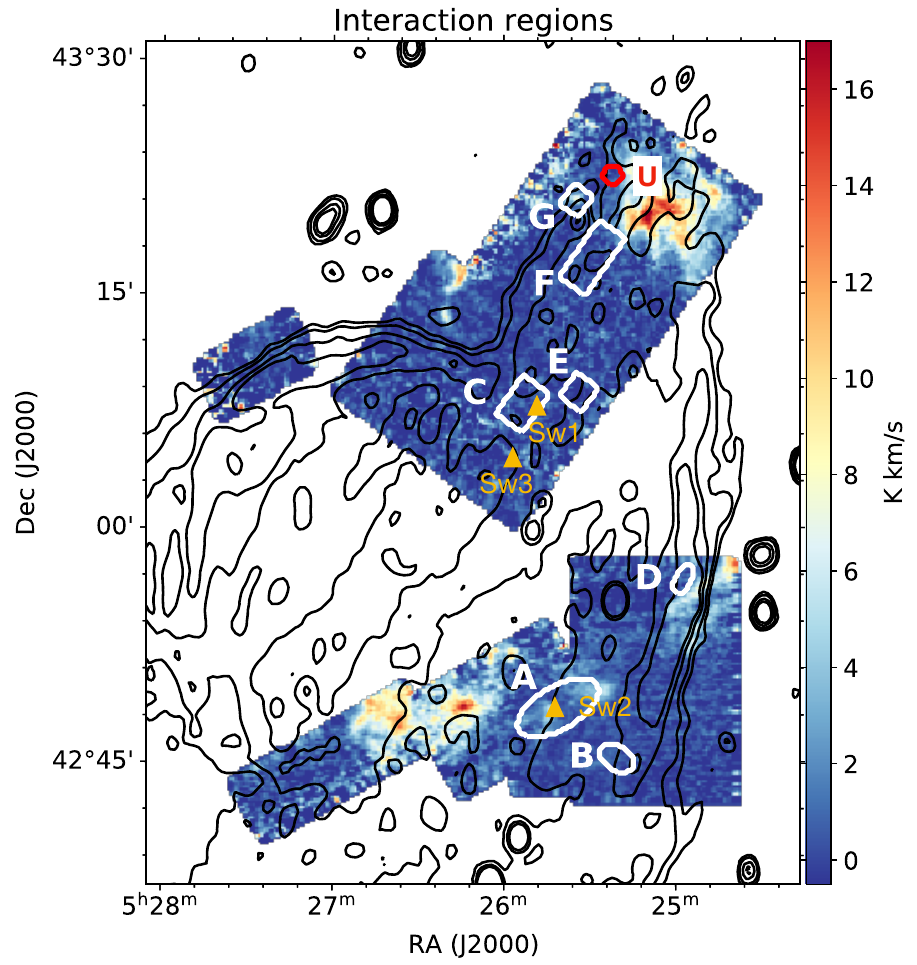}
\caption{Regions of the disturbed molecular material around VRO 42.05.01. Location of the regions described in table \ref{table:vro}, overlaid on the footprint of IRAM observations, and showing the 1420~MHz contours from the CGPS. Their line profiles are shown in Fig. \ref{fig:inverted_lines}. 
}
\label{fig:inverted_lines_locations}
\end{figure}

The regions labelled in Fig.~\ref{fig:line_intensities} as A and B are easy to discern by eye, from the relatively brighter \twcoto\ emission as compared to the \twcooz\ emission (region A) and the broad velocity dispersions (regions A and B). Since the \twcoto\ emission is often faint, and in the map of velocity dispersions, lower, but still broadened, values of the best-fit standard deviation for the fitted Gaussian can be confused with noise, we further looked for regions of interaction by making a grid of rectangular regions covering the remnant. Since we had not taken this additional step in \cite{arias19}, we also included those data in our analysis. We identified seven regions (A through F, see Figs. \ref{fig:inverted_lines} and \ref{fig:inverted_lines_locations}, and table \ref{table:vro}) that show possible signatures of interaction, based on their having a \twcoto\ to \twcooz\ ratio larger than one. For comparison, we show in the bottom, right corner of Fig. \ref{fig:line_intensities} the line profile of region U, which we picked to show the properties of an unperturbed molecular line: a velocity dispersion of 1.0~km~s$^{-1}$ and a \twcoto\ to \twcooz\ ratio of 0.65.

The line profiles for the regions of interaction are shown in Fig. \ref{fig:inverted_lines}, their locations in Fig. \ref{fig:inverted_lines_locations}, and their properties are summarised in table \ref{table:vro}. The three regions that most clearly show evidence that the SNR shock is interacting with the molecular gas are regions A, B, and D. In all three cases, the lines are broadened (full-width at half-maxima, FWHM, of $>10$~km~s$^{-1}$) and the ratio of the intensity peaks between the \twcoto\ and \twcooz\ transitions is approximately 2. For the remaining lines, the evidence is weaker: the velocity dispersions are narrower ($<4.5$~km~s$^{-1}$), and the \twcoto\ to \twcooz\ ratios range between $1.1-1.5$. The broad range in central velocities (see table \ref{table:vro}) suggests that it is unlikely that all of these features are related to a SNR and molecular cloud interaction.

The reason underpinning the use of inverted $^{12}$CO line ratios as diagnostic of SNR and molecular cloud interaction has to do with the temperature and/or density conditions in the molecular cloud requires for the higher-$J$ transitions to be more populated than the lower-$J$ transitions. 
We can derive some properties of the gas by performing radiative modelling with the statistical equilibrium radiative transfer code RADEX \citep{radex}.

\begin{figure}
\centering
\includegraphics[width=\columnwidth]{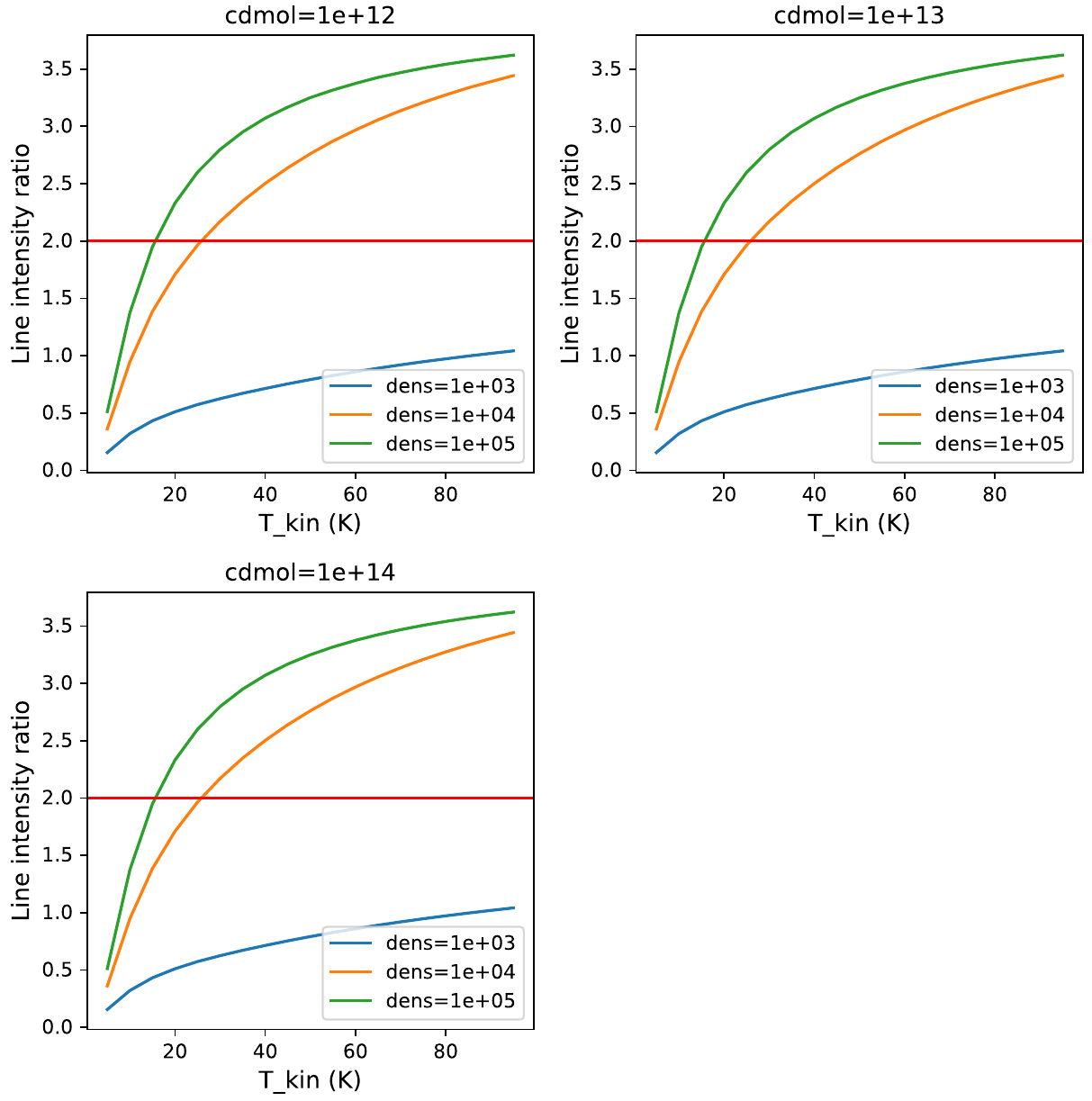}
\caption{Results from the radiative modelling of the \twcoto\ and \twcooz\ transitions with RADEX. The ratio between the \twcoto\ and \twcooz\ line peaks ratio is plotted as a function of kinetic temperature (\texttt{T\_kin}) for different values of H$_2$ density (\texttt{dens}, in cm$^{-3}$) and column density (\texttt{cdmol}, in cm$^{-2}$). A horizontal line for ratio equals to 2 indicates the conditions of the molecular gas in the regions A, B, and D around VRO 42.05.01.
}
\label{fig:radex}
\end{figure}

Figure \ref{fig:radex} shows the results from the RADEX modelling. The horizontal line indicates the parameter space for which to the average line intensity peaks ratio is 2, as is the case for regions A, B, and D. For all combinations of parameters explored, the observed ratio can only be explained if the molecular material has a kinetic temperature (\texttt{T\_kin}) higher than 10~K. The modelling is quite independent of column density (the variable named \texttt{cdmol} in Fig. \ref{fig:radex}; column density affects the total intensity of the lines, but not their ratio) and favours either rather high molecular densities ($10^5$~cm$^{-3}$) and $\sim20$~K, or lower densities ($10^4$~cm$^{-3}$) but warmer gas conditions ($\sim30$~K).

\subsection{Further evidence of swept up molecular material around VRO 42.05.01}

\begin{figure*}
\centering
\includegraphics[width=0.7\textwidth]{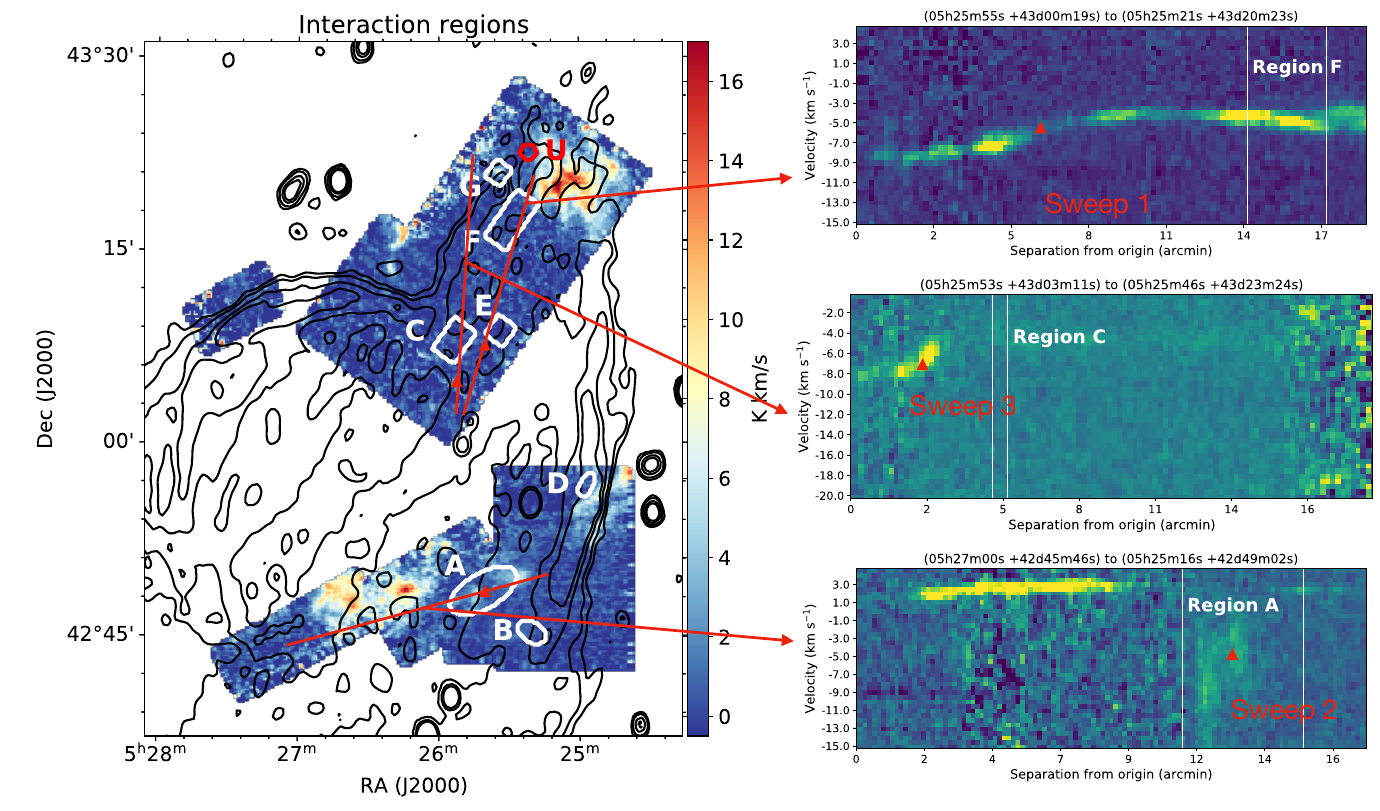}
\caption{Position-velocity diagrams for three cuts along the VRO 42.05.01 \twcooz\ data cube. Each of the cuts is shown as a red line over the integrated molecular emission map, and the locations of the sweeps, where the bright molecular emission smoothly changes velocity channels, are shown as red triangles both in the map and in the position-velocity diagrams. Only Sweep 2 is related to a shock heated feature (Region A). Sweep 1 was identified in \cite{arias19}.
}
\label{fig:pos_vel}
\end{figure*}

In \cite{arias19} we found a velocity gradient in a particular molecular cloud region (labelled \lq Sweep 1\rq\ in Fig. \ref{fig:pos_vel}) where the central velocity of the emission moved from $-9.0$~km~s$^{-1}$ to $-4.5$~km~s$^{-1}$ over a length of $\sim10'$.
We argued that this gradient is the result of the molecular cloud's interaction with VRO~42.05.01's progenitor stellar wind. We searched for similar gradients in the new data, and re-examined the old data for velocity gradients we missed due to the low brightness temperature of the emission. 
Fig. \ref{fig:pos_vel} shows position-velocity diagrams for three cuts across the molecular emission. Sweep 1 corresponds to the region identified in \cite{arias19}. Sweep 3 shows a steeper gradient, with the central velocity of the emission changing from $-8.0$~km~s$^{-1}$ to $-4.5$~km~s$^{-1}$ over $\sim2'$. The position-velocity plots in Sweeps 1 and 2 both have narrow velocity dispersions in the emitting material (seen as vertical cuts across the position-velocity plot for any given value of the x-axis), and are close to each other in velocity-space, and in position (see left-hand side plot of Fig. \ref{fig:pos_vel}), implying they could be related, and both be imprints of the same gust of stellar wind on the molecular gas.

Whatever is pushing the gas and causing Sweeps 1 and 3 is not affecting its velocity dispersion; it is mechanically displacing the gas, but not heating it to the point that it is visible as broadened lines. This means that the material has either cooled to its quiescent temperature after a shock passage, or was never heated, which is consistent with our initial suggestion \citep{arias19} that the molecular gas is not swept up by the SNR shock, but rather by a stellar wind of the progenitor star. Cooling rates for shocked molecular clouds depend on their particular temperatures, composition and densities \citep{dalgarno72}; it is possible that a stellar wind shock heated the material and swept it up, and that material has since cooled, but its swept up morphology remains visible.

The bottom plot in Fig. \ref{fig:pos_vel}, corresponding to Sweep 2,  is different: it also shows a sweep in position-velocity space, but this time the sweep shows a large velocity dispersion $\Delta v\approx10$~km~s$^{-1}$ (the bright emission at positive velocities is likely a foreground component). The location of the sweep corresponds with Region A, which we had previously identified as a region where the SNR shock is interacting with the molecular gas from its high \twcoto\ to \twcooz\ line ratio and broad line FWHM. A possibility for this structure in position-velocity space is that the molecular material was first swept by a stellar wind, just like Sweeps 1 and 3, and that the SNR shock is now encountering and heating this swept-up gas.

\subsection{Location of the SNR with respect to the molecular features}

In table \ref{table:vro} we list regions A through G, and sweeps 1 through 3, in descending order of central velocity. If all these features were related to the VRO 42.05.01 SNR shock, or winds from its progenitor star, then the remnant would be associated with features at LSR velocities of $-22.5$~km~s$^{-1}$ to $-5.0$~km~s$^{-1}$, which is unlikely, suggesting that some of these features might be unrelated to the SNR and be the result of a different source of heating.

As discussed previously, regions A, B, and D show the clearest evidence for shock heating, presumably from the SNR shock. Their central velocity values are $-7.5$~km~s$^{-1}$, $-10.0$~km~s$^{-1}$, and $-14.0$~km~s$^{-1}$, respectively. However, the central velocities of broadened lines are often not the same as their systemic velocities, especially when the shock direction is not perpendicular to the line of sight. This evidence of an interaction between the SNR shock and the molecular cloud is therefore still consistent with a SNR LSR velocity of $-6.0$~km~s$^{-1}$ and a 1~kpc distance \citep{reid14,arias19}. 

If all, or some, of the features we noted in table \ref{table:vro} other than features A, B, and D (spanning a velocity range of $-22.5$~km~s$^{-1}$ to $-5.0$~km~s$^{-1}$) are in fact related to VRO~42.05.01, and the SNR shock is encountering the molecular gas in regions A, B, and D, then some, if not all, of regions C and E -- G are shocked by something of larger radius than the SNR shock ---possibly a stellar wind. This would be consistent with their line properties, indicative of a less strong shock than that encountering A, B, and D. This interpretation requires further observations and modelling; we simply note it here because, for the sake of completeness, we report all instances of heated gas --as evidenced by a \twcoto\ to \twcooz\ ratio larger than one-- in our observations in table \ref{table:vro}.


\section{Results -- G 350.0-2.0}


\begin{figure*}
\includegraphics[width=\textwidth]{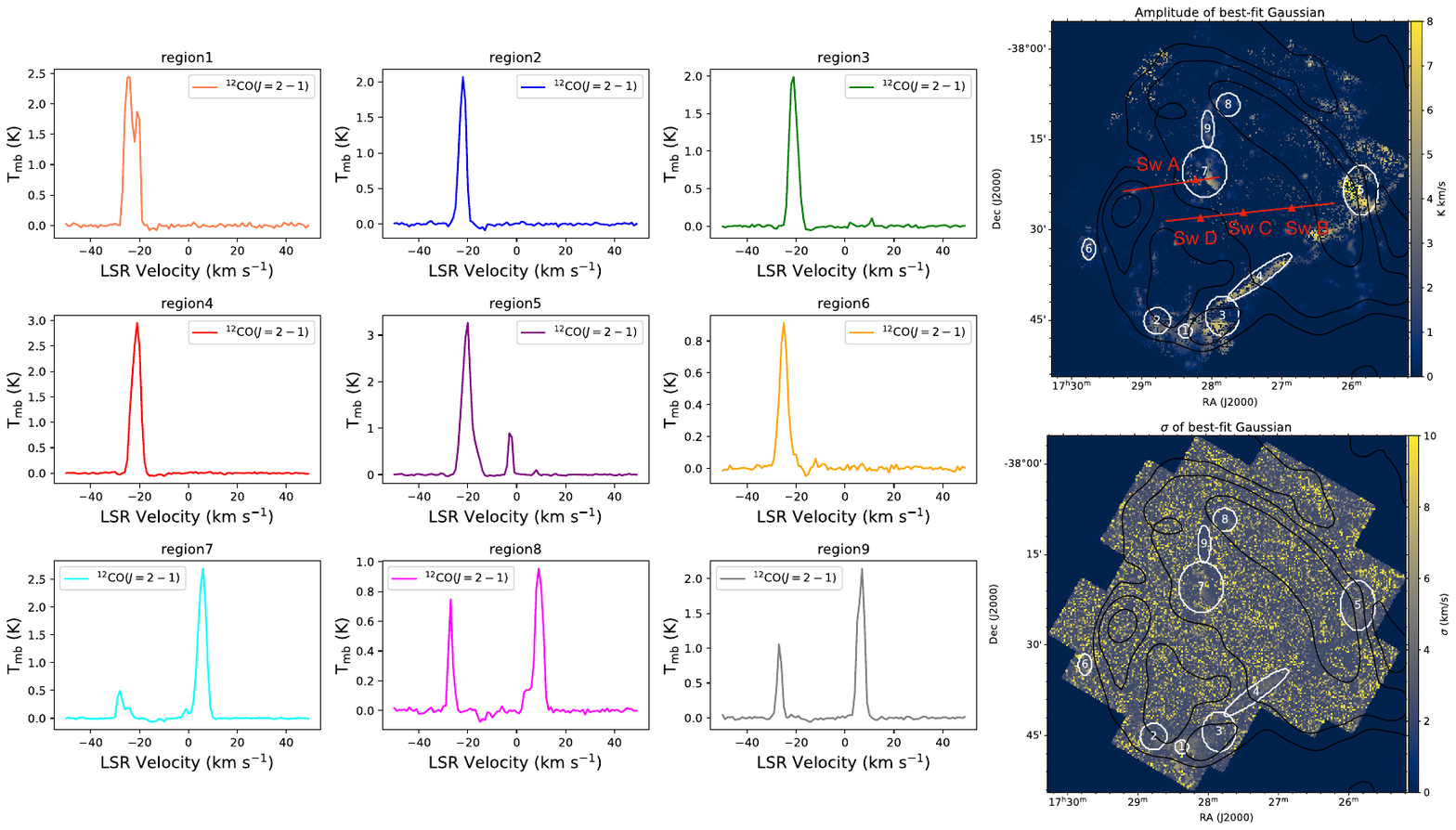}
\caption{Properties of the \twcoto\ emission towards G 350.0-2.0. \textit{Left.} Emission lines for a selection of regions (labelled in the right-hand-side maps) in G 350.0-2.0. The asymmetrical wings in regions 5 and 6 hint at the possibility of interaction, but there seems to be no evidence of shocks from broadened lines in any of the regions. \textit{Right.} Best-fit amplitude (top) and standard deviation (bottom) after fitting a Gaussian function to each pixel in the data cube. The lines in red on the amplitude plot correspond to the two position-velocity slices shown in Fig. \ref{fig:G350_posvel}. The overlaid contours are from the 88~MHz GLEAM image of G 350.0-2.0 \citep{hurley-walker19}.
}
\label{fig:G350gaussians}
\end{figure*}

\subsection{Spectral index map and low-frequency steepening}

The spectral index map of G 350.0-2.0 (Fig. \ref{fig:g30_spx}) is similar to that of VRO 42.05.01 \cite[see][fig. 4]{arias19b}. We observe in both cases that the spectral index in the bright regions is steeper than in the faint regions, and that it is flatter than $\alpha = -0.5$. In general, mixed morphology SNRs have relatively flat spectral indices \citep{vink12}.

In addition to the 200~MHz maps, the GLEAM data products also consist of FITS files at 88~MHz. After subtracting the background, we measured the integrated flux density of G~350.0-2.0 to be $S_\mathrm{88~MHz} = 88.3 \pm 8.8$~Jy. We do not present a spectral index map between 200~MHz and 88~MHz because the background in both GLEAM maps shows a gradient across the surface of the remnant that results in an artificially steeper shell spectral index value and artificially flatter wing. The flux density measurement alone, though, indicates that the spectrum of G 350.0-2.0 steepens at low radio frequencies, with $\alpha \approx -0.7$ between 88~MHz and 200~MHz, versus $\alpha \approx -0.4$ at higher frequencies. We observed this behaviour in VRO 42.05.01 \citep{arias19b}, and we proposed that in high compression ratio shocks, the low-frequency electrons penetrate deeper into the downstream plasma than their higher frequency counterparts, thereby experiencing larger compression ratios, and hence resulting in a steeper low-frequency spectrum. We were able to make the argument that VRO 42.05.01's shock has high compression ratios because it shows abundant optical line emission indicative of radiative shocks. However, we do not know whether G 350.0-2.0 has a radiative shock ---\cite{stupar11} detected very little filamentary emission towards this region, although it is worth noting that towards the inner Galaxy optical emission is much more likely to be absorbed than in the direction of the Galactic anticentre, where VRO 42.05.01 is located.

\subsection{Molecular line features}

\begin{table*}[]
\centering
\begin{tabular}{cccccc}
Region name &
  Central RA &
  Central Dec &
  \begin{tabular}[c]{@{}c@{}}Central velocity\\ (km~s$^{-1}$)\end{tabular} &
  \begin{tabular}[c]{@{}c@{}}FWMH\\ (km~s$^{-1}$)\end{tabular} &
  \begin{tabular}[c]{@{}c@{}}Signal-to-noise ratio\\ of \twcoto\end{tabular} \\ \hline
Region 1 & 17:28:22 & -38:46:52 & -24.0 & 4.0 & 99.0  \\
Region 2 & 17:28:46 & -38:45:09 & -22.0 & 2.0 & 108.0 \\
Region 3 & 17:27:50 & -38:44:20 & -21.0 & 3.0 & 100.5 \\
Region 4 & 17:27:19 & -38:37:52 & -21.0 & 3.0 & 187.3 \\
Region 5 & 17:25:52 & -38:23:28 & -20.0 & 3.0 & 195.1 \\
Region 6 & 17:29:44 & -38:33:15 & -25.0 & 2.0 & 74.5  \\
Region 7 & 17:28:05 & 38:20:25  & 6.0   & 3.0 & 345.5 \\
Region 8 & 17:27:44 & -38:09:17 & 9.0   & 3.0 & 92.4  \\
Region 9 & 17:28:02 & -38:13:17 & 7.0   & 3.0 & 151.1 \\
Sweep A & 17:28:12 & -38:22:10 &  4.0  &  & \\
Sweep B & 17:26:54 & -38:28:19 &  2.0  &  & \\
Sweep C & 17:27:37 & -38:28:33 &  4.0  &  & \\
Sweep D & 17:28:11 & -38:29:08 &  9.0  &  & \\
\end{tabular}
\caption{Properties of selected regions towards G~350.0-2.0. Regions 1 through 9 correspond to the emission regions whose line profiles are displayed in Fig. \ref{fig:G350gaussians}, and Sweeps A through D are shown in Fig. \ref{fig:G350_posvel}.
}
\label{table:g350}
\end{table*}

\begin{figure}
\includegraphics[width=\columnwidth]{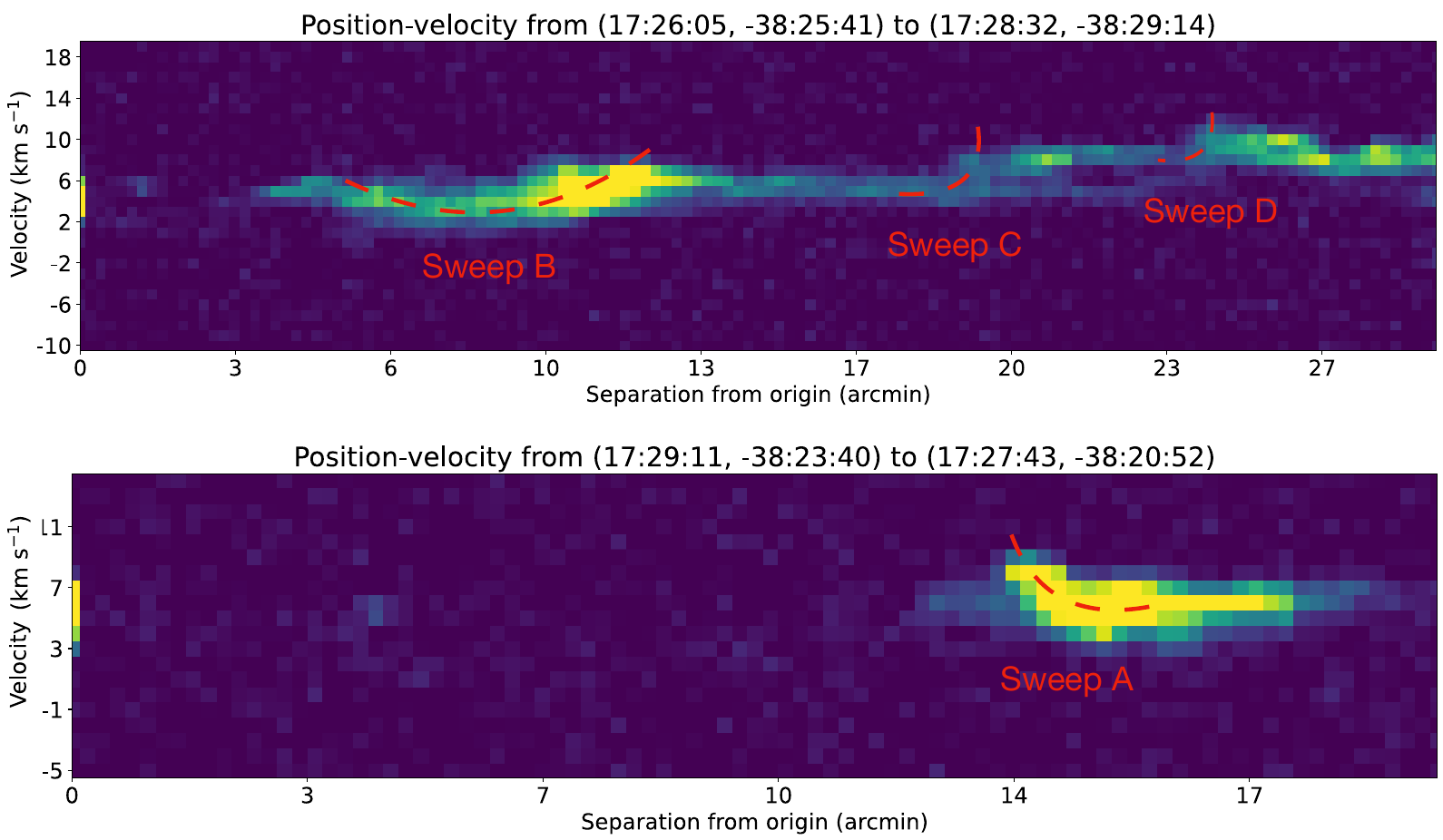}
\caption{Position-velocity diagrams around two lines-of-sight in G~350.0-2.0. The lines-of-sight in question are shown as red lines in the amplitude image in Fig. \ref{fig:G350gaussians}, with the location of the midpoints of the arcs indicated with dashed red lines in these position-velocity diagrams shown as red triangles.
}
\label{fig:G350_posvel}
\end{figure}

Figure \ref{fig:footprint}, bottom panel, shows the total velocity-integrated emission per pixel for the area around G 350.0-2.0. Much like for VRO 42.05.01, the molecular environment is patchy and uneven, and does not have any clear morphological similarities with the SNR. There are two major emission components, one at negative velocities (from $-28$~km~s${^{-1}}$ to $-16$~km~s${^{-1}}$), and one around zero velocity ($-3$~km~s$^{-1}$ to $+11$~km~s$^{-1}$), but the transition between them is rather continuous, with no truly emissionless channels.

The APEX observations presented here only cover a single CO transition, \twcoto. This means that we cannot look for regions where the molecular material is shocked via inverted line ratios. Instead, we looked for broadened velocities and for swept up material showing as arcs in position-velocity space. 

Figure \ref{fig:G350gaussians} shows the amplitude and standard deviation results of fitting a Gaussian function to each pixel in the data cube. We also show the integrated line profiles for the regions showing the bulk of the emission (labelled regions 1 through 9). The line properties are shown in table \ref{table:g350}. We examined the map of the best-fit standard deviation, but did not find any lines broader than $3\pm1$~km~s$^{-1}$. Region 1 appears to be broadened, but in the plotted flux per channel for the region (see top-left plot in Fig. \ref{fig:G350gaussians}) we can appreciate that there are two blending narrow components to this seemingly broadened line. The spectral lines in the plots of Fig. \ref{fig:G350gaussians} have an average FWHM of 3~km~s$^{-1}$. The only hint at an interaction is that regions 5, 6, and 7 show asymmetrical wings in the line component at around $-20$~km~s$^{-1}$, but this could well be line blending; it constitutes weak evidence of a possible interaction, and more data are required to assess it.

In terms of swept up material in position-velocity space, the evidence, too, remains inconclusive. Figure \ref{fig:G350_posvel} plots the most promising \lq arcs\rq\ that we found in the data in the two directions shown in red in Fig. \ref{fig:G350gaussians}, top right. The positions of the sweeps are listed in table \ref{table:g350}. Along both lines the observed $\Delta v$ are relatively small compared to the resolution of the observations (compare to the sweeps seen in VRO 42.05.01 in Fig. \ref{fig:pos_vel}); these are for now no more than hints of the molecular material having been swept up.

From our APEX observations, we cannot confirm any interaction between the SNR and its neighbouring molecular material. Although we only have access to one molecular transition, the mapped area is large, and shows abundant molecular emission, yet we find no indications that the molecules are heated. Our data, while they can neither confirm nor reject interaction between the SNR and the molecular clouds, appears to support lack of interaction. The dense molecular environment does not appear to be responsible for G 350.0-2.0's large-scale, shell and wing morphology.

\section{Discussion}

\begin{figure}
\centering
\includegraphics[width=\columnwidth]{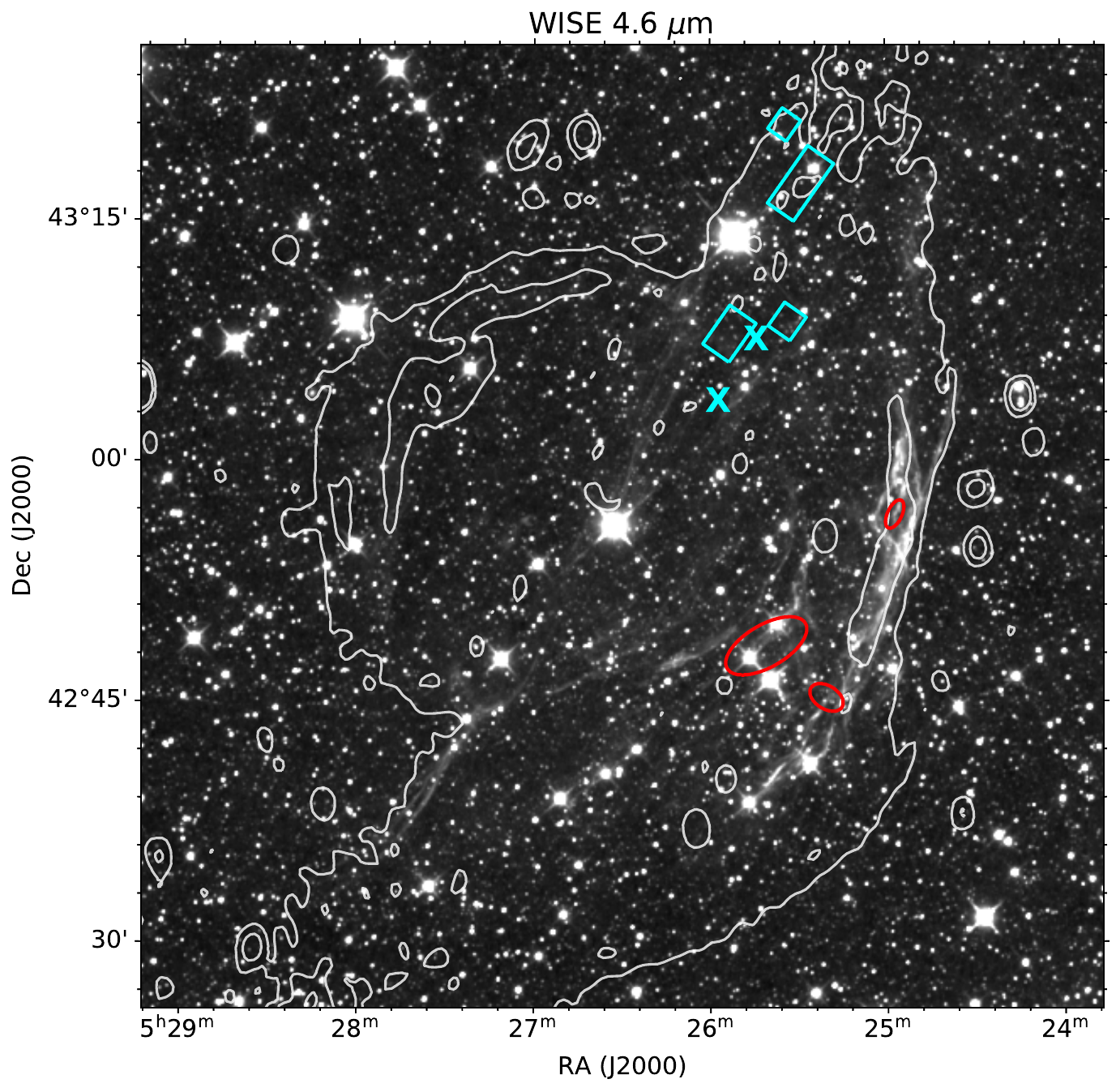}
\includegraphics[width=\columnwidth]{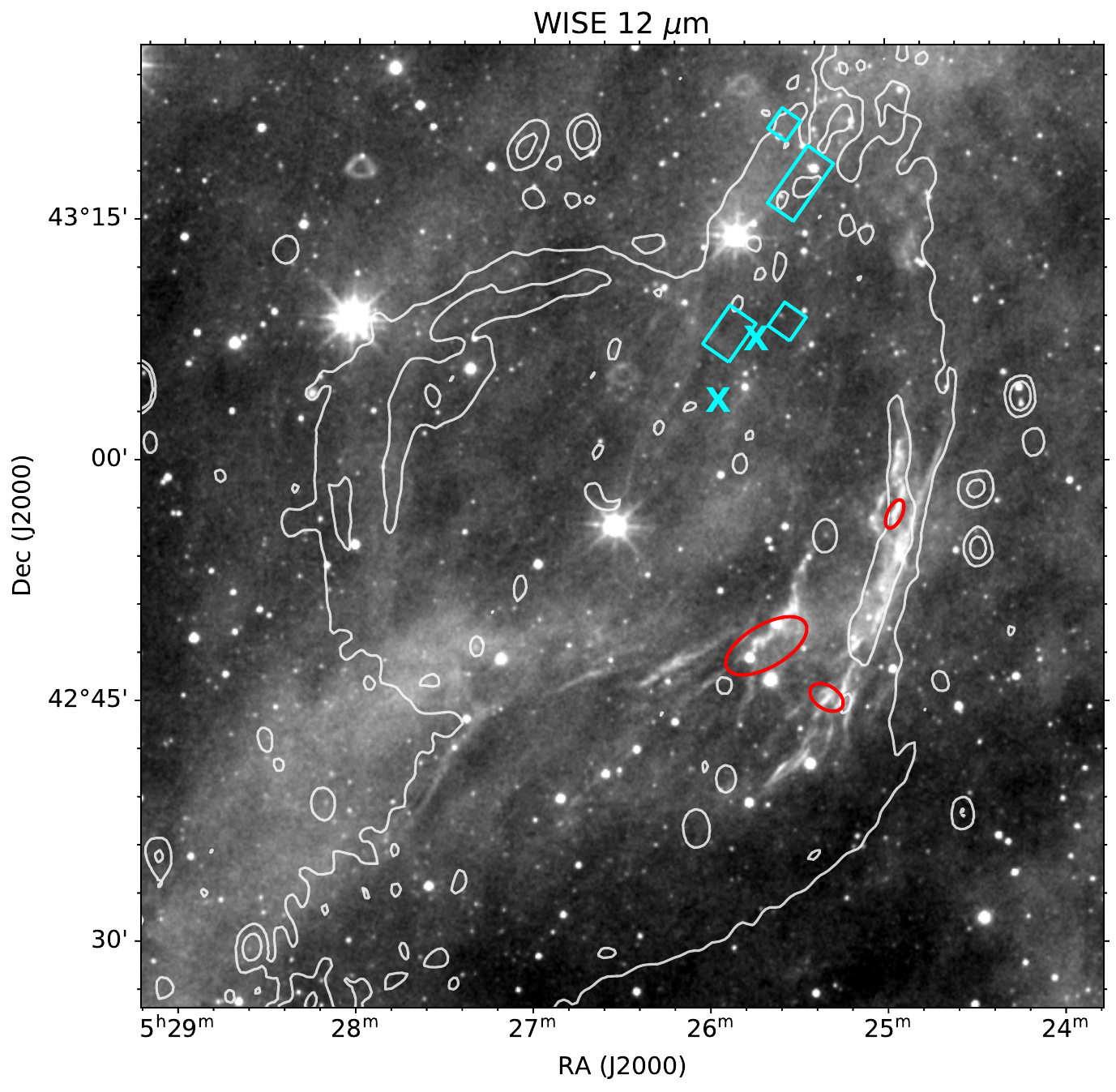}
\caption{Infrared emission from VRO 42.05.01 at 4.6~$\mu$m (top) and 12~$\mu$m (bottom), with the regions identified in Fig. \ref{fig:inverted_lines_locations} overlaid. The sweeps in position-velocity space from Fig. \ref{fig:pos_vel} are indicated with an X. The features in red are those we believe are due to the SNR shock, and (some or all) those in cyan could be due to a pre-SN stellar wind.
We note that the regions of swept-up mass occur along the shell of VRO~42.05.01. Both maps are taken from the \textit{WISE} archive \citep{wright10}.
}
\label{fig:vrowise}
\end{figure}

\begin{figure}
\includegraphics[width=\columnwidth]{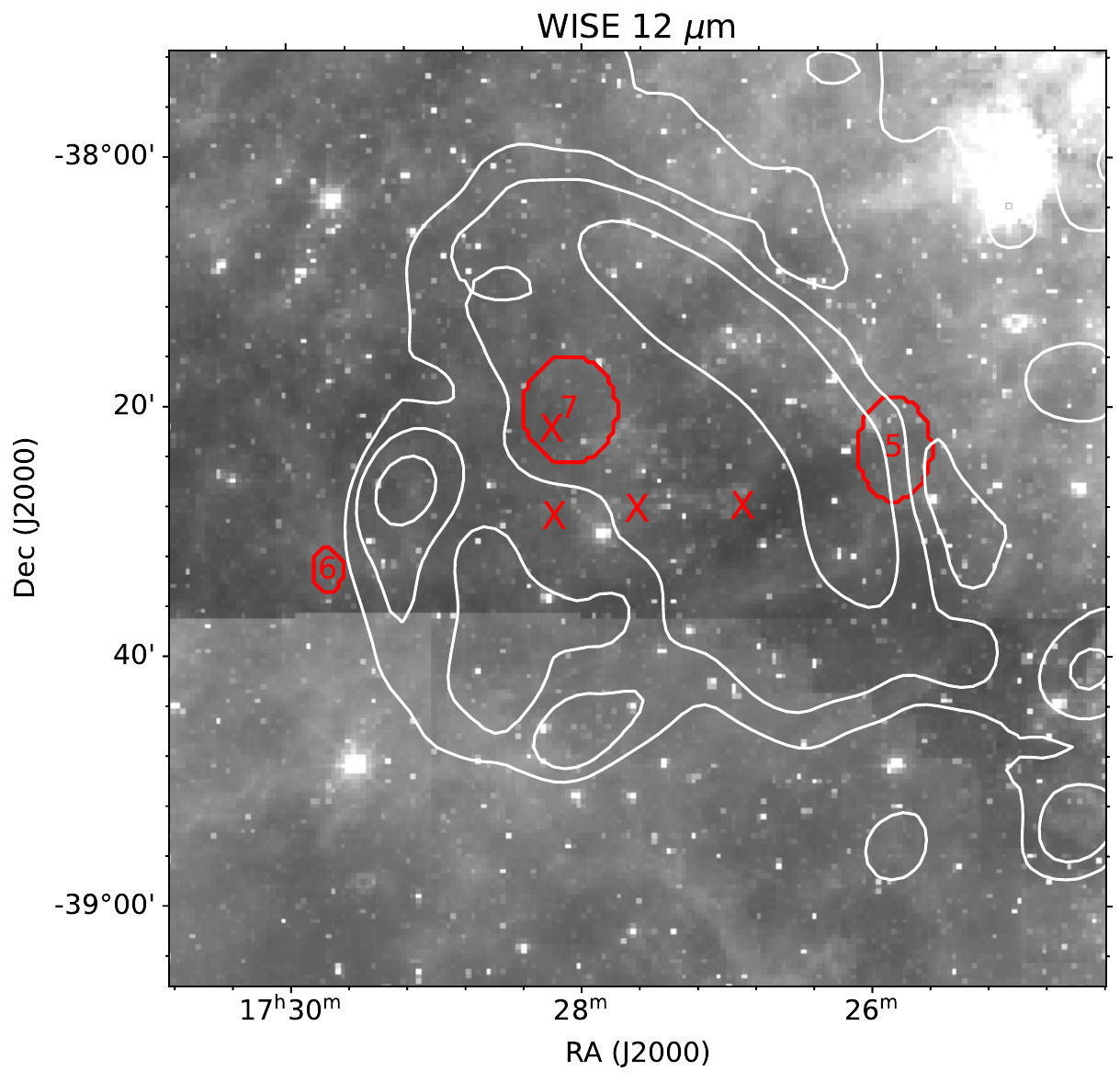}
\caption{G 350.0-0.2 at 12~$\mu$m as seen with \textit{WISE} \citep{wright10}. The radio contours are from the 88~MHz GLEAM image of G 350.0-2.0 \citep{hurley-walker19}. The regions labelled in red correspond to regions 6, 7, and 8 in Fig. \ref{fig:G350gaussians}, and the sweeps in position-velocity space are labelled with an X.
}
\label{fig:G350_wise}
\end{figure}

\subsection{Infrared environment of VRO 42.05.01 and G 350.0-2.0}
\label{sec:ir}

Following the discussion in \cite{arias19} we show the infrared emission of VRO 42.05.01 at 4.6~$\mu$m and  12~$\mu$m from the Wide-field Infrared Survey Explorer \cite[\textit{WISE,}][]{wright10} in Fig.~\ref{fig:vrowise}. We show these two bands because they present the most features towards VRO 42.05.01 (see that in fig. 12 of \cite{arias19} all the --relatively few-- features visible in the \textit{WISE} 22~$\mu$m map are visible in the 12~$\mu$m map). 

We had previously identified several possible regions for interaction \cite[see fig. 13 in][]{arias19}: a thick, \lq scarf\rq-like band of dust that seems to wrap around the VRO 42.05.01 shell (the only one of the features that is visible at 22~$\mu$m), a bright synchrotron filament in the west of the wing (also bright in the \textit{WISE} bands at 3.4~$\mu$m, 4.6~$\mu$m and 12~$\mu$m and in optical H$\alpha$ and [S {\sc ii}] lines), and a series of filaments internal to the wing that trace the continuation of the semicircle of the small shell (visible at 4.6~$\mu$m and 12~$\mu$m). 
We also discussed the neutral hydrogen environment of VRO 42.05.01, and found what appears to be a cavity at the location of VRO 42.05.01 at $-6$~km~s$^{-1}$. In H {\sc i}, the denser features are North of the shell, at $-6$~km~s$^{-1}$, and south of the wing, at $-11$~km~s$^{-1}$ \cite[see Fig. 11 in][]{arias19b}. There is no clear overlap between the infrared and molecular features, and the H {\sc i} gas.

The regions of interaction of the VRO 42.05.01 SNR shock with the molecular gas (regions A, B, and D, in red in Fig. \ref{fig:vrowise}) show corresponding infrared emission at 4.6~$\mu$m and 12~$\mu$m. Region D overlaps with the southern edge of the bright synchrotron/IR/optical filament, but the inverted ratios are confined to a very small part of this feature, in spite of a large fraction of the filament being covered in the CO observations and some of it showing molecular line emission. On the other hand, features C, E, F, G, and Sweeps 1 and 3 (all shown in cyan in Fig. \ref{fig:vrowise}) follow the scarf of emission at $12~\mu$m and $22~\mu$m that wraps around the VRO 42.05.01 shell. The scarf is the only feature that seems to have a large-scale correspondence with the shell and wing morphology of VRO 42.05.01. The location of the features in cyan, which we have tentatively attributed to a stellar wind, seems to be on the plane (parallel to the Galaxy) separating the shell and the wing, consistent with an equatorially enhanced wind.

We show the $12~\mu$m \textit{WISE} emission of G 350.0-2.0 in Fig.~\ref{fig:G350_wise}, with regions 5, 6, and 7 in Fig. \ref{fig:G350gaussians} labelled (these are the most likely regions for interaction that we can find in our APEX data, as they show asymmetrical wings), and the swept-up features in position-velocity space indicated with an X. The infrared emission in the G 350.0-2.0 area is quite different from that of VRO 42.05.01 --in particular, there is no infrared emission tracing the radio emission, and there is no apparent density discontinuity between the shell and the wing. Regions 5, 6, and 7 are plausible regions for interaction, in the sense that they are located where the SNR shock is encountering the ISM (regions 5 and 6), and between the shell and the wing (region 7), but the \textit{WISE} maps gives no further hint of interaction. The possible swept-up features labelled A, C, and D are located in a similar disposition to the swept-up features of VRO 42.05.01, roughly along the vertical line separating the shell and the wing.
Unlike for VRO 42.05.01, the infrared environment of G 350.0-2.0 shows no morphological similarities or large-scale features enveloping the radio emission. 

In addition to the infrared, these two SNRs also look quite different in the optical; however, the relative lack of optical observations of G 350.0-2.0 precludes us from drawing strong conclusions from this difference. \cite{stupar11} observed G 350.0-2.0 using the he Anglo-Australian Observatory/United Kingdom Schmidt Telescope (AAO/UKST) H$\alpha$ survey. The image they use covers the H$\alpha$ and the [N~\textsc{ii}] 6548 and 6584 emission lines, but we do not have any information about the SNR in other important diagnostic lines, such as [S~\textsc{ii}], [O~\textsc{iii}] and H$\beta$. To our knowledge, there are no spectroscopic observations of this source either. It is possible that the source has [S~\textsc{ii}] and [O~\textsc{iii}] radiative emission that we have not observed, and so we cannot meaningfully discuss the optical differences between the two sources.

\subsection{The relationship between the shell and wing sources and their environments}

VRO 42.05.01 and G 350.0-2.0 look strikingly similar in their radio emission, but their environments appear to have little in common. Similarities in the sources are: their mixed morphology nature, their shell and wing shape, their large-scale orientation with respect to the Galaxy (the boundaries between shell and wing are parallel to the Galactic plane, the wing points in the direction of the Galaxy), and their flat integrated radio spectral index that steepens at low radio frequencies. The molecular environment of VRO 42.05.01 is relatively sparse, but enhanced CO line ratios and arcs of emission in position-velocity space indicate that the SNR shock (or perhaps a progenitor wind) is interacting with the molecular environment. On the other hand G 350.0-2.0 shows no line broadening, arguing against an interaction with its molecular environment (however, this needs to be confirmed with observations at a different transition to obtain a more firm conclusion). A further difference is their infrared environment. Unlike for VRO 42.05.01, the infrared emission of G 350.0-2.0 does not show any small-scale or large-scale correspondence with its radio morphology.

Our aim in studying these sources jointly is to understand the role their environment plays in developing their unusual shell and wing shape. If these sources are so similar, and their strange shell and wing morphology is a result of interactions with their environment, then we would expect to see similar large-scale ambient environmental features. This is contrary to what we observe. 

As we mentioned earlier, the series of papers by \cite{pineault85,pineault87} and \cite{landecker89} proposed that a density discontinuity in VRO 42.05.01's surrounding ISM is responsible for its shell and wing morphology. However, our analysis of the molecular, and neutral hydrogen environments of these sources shows no evidence for such a density discontinuity (the band of emission we refer to as the scarf, visible at $12~\mu$m and $22~\mu$m, is the only large-scale feature with a morphological correspondence with the SNR). Moreover, in our observations (with the caveat that we have not fully covered the SNR shell) the dense molecular gas is in the wing, whereas if the shape of VRO 42.05.01 were due to a density discontinuity, we would expect the wing to be in the low density side of the discontinuity and the shell in the high density side. 

Given that both VRO~42.05.01 and G~350.0-2.0 share a shell and wing morphology, it seems more likely that they share a common evolutionary history of their progenitor stars than that they both coincidentally sit in an ISM density discontinuity that we are unable to detect.
In our earlier work on VRO~42.05.01 \citep{arias19,arias19b} we suggested that a mass-losing, supersonically moving progenitor star that exploded could presumably result in a shell and wing morphology, with the bow shock of the moving star accounting for the wing. Hydrodynamic modelling has shown that massive runaway stars result in asymmetric SNRs \citep{meyer15}, with the size and velocity of the progenitor star greatly influencing the SNR morphology \citep{meyer23}.

In \cite{Chiotellis2019} we performed hydrodynamic simulations of the explosion of a Wolf Rayet runaway star, and were able to reproduce the kinematics and morphology of VRO~42.05.01. We invoked a Wolf-Rayet progenitor because we took the distance to VRO 42.05.01 to be 4.5~kpc \cite[as proposed by][]{landecker89}, which required a very large wind cavity to account for its observed angular size \cite[we later showed in][that the distance to VRO 42.05.01 is in fact much less than that, and  proposed a value of $1.0\pm0.4$~kpc]{arias19}. 
A runaway progenitor scenario was not enough to account for the shell and wing morphology, and we further invoked that the stellar wind of the progenitor star is not spherically symmetric, but rather bipolar and equatorially confined \citep{Chiotellis2019}. This is consistent with the point presented in Fig. \ref{fig:vrowise} that the molecular features that could be due to a stellar wind (in cyan) are roughly along the equator of the SNR (the plane separating the shell and the wing). 

A further point to note here is that if the wing was indeed formed by the bow shock of a runaway progenitor, then the common orientation of the two remnants (both wings point towards the Galactic plane) implies that the progenitor stars of both remnants were moving vertically towards the Galactic plane. Runaway stars infalling into the plane are not uncommon \cite[e.g. AB Cru, $\theta$ Ara,
HD 125288, 67 Oph, HD 161695;][]{maizapellaniz18}, and interestingly, the majority of runaway stars are also fast rotators, as expected from
the supernova scenario of runaway stars. This further supports a
bipolar and equatorially confined stellar wind of the shell and wing
progenitors as responsible for their morphology.

Perhaps more importantly, the two sources are mixed morphology SNRs. The development of a mixed morphology has traditionally \cite[e.g.][]{cox99,chen08} been proposed to be due to interaction with a dense environment. For instance, \cite{ustamujic21} were able to replicate the properties of IC~443, a SNR in a notoriously inhomogeneous medium, as a result of the interaction of the SNR shock with its environment, including the centrally peaked X-ray morphology (i.e. the mixed morphology). 

However, for the two shell and wing sources under consideration here, there does not seem to be enough of a dense medium around them to account for their mixed morphology in the same way as for IC~443. 
If we consider a bow shock due to a mass-losing, supersonically moving progenitor, then we can invoke the reflected shock scenario for the development of a mixed morphology as proposed by \cite{chen08}. In this case, the structure of the circumstellar environment, once the star explodes, could provide the cavity for the blast wave to be reflected and result in a centrally-peaked X-ray morphology surrounded by a radio shell. Detailed hydrodynamical simulations would be required to test the validity of this hypothesis.

If we adopt this scenario, then the wind bubble size should be comparable to the current SNR radius. Using the linear relation of \cite{chen13} we can roughly estimate the progenitor mass of the two SNRs by their mean radius. This gives a main sequence mass of 16~\msun\ for G 350.0-2.0 ($R\sim 10$~pc) and of 14~\msun\ ($R\sim 8.4$~pc) for VRO 42.05.01. The similar main sequence mass estimation for these two sources further supports that the two stars could share a common stellar evolutionary history.

\subsection{A comparison of VRO 42.05.01 and G 350.0-2.0 with other mixed morphology SNRs}

The first clear-cut example of the interaction of a SNR with a molecular cloud  was the case of IC~443, which shows broadened molecular lines of a variety of species \citep{denoyer79,vandishoeck93}, excited molecular hydrogen \citep{treffers79}, and OH maser emission \citep{claussen97}. \citet{jiang10} compile a list of SNRs interacting with molecular clouds and the existing evidence for that interaction; the SNRs with the most compelling evidence tend to be of the mixed morphology class (e.g. IC 443, W44, W28, W51, 3C 391).

IC~443, 3C 391, W44 and W28 are all associated with giant molecular clouds (GMCs): W44 and W28 are thought to be expanding inside their parent GMC \citep{reach05}, whereas 3C~391 and IC~443 have been proposed to be expanding near the edge of a GMC \citep{reach99,cornett77}. We observe no such large-scale molecular emission towards VRO 42.05.01; in fact, the molecular material appears to be clumpy and small in size as compared to the SNR, and within it, the areas of shock interaction are even smaller (for instance, the semimajor axis of region D subtends $56''$, or 0.27~pc at a distance of 1~kpc). Some SNRs that are interacting with molecular clouds show indentations coincident with the location of the molecular material \cite[e.g. see figs. 11 and 1 in][for W44 and IC~443, respectively]{seta04,bykov08} that suggest that the dense ISM can account for large-scale morphological features of the SNR; again, in the case of VRO 42.05.01, we find no such indentations. Interestingly, \citet{reach99} observe a large area over 3C 391 in CO where there is molecular emission, but find that only a small region ($<0.6$~pc) shows shocked line features --much like we find in VRO 42.05.01-- suggesting perhaps that these signatures of shocked interactions (broadened lines) are either short-lived, or require density conditions that are not found everywhere in the molecular cloud.

The velocity dispersions of $\gtrsim10$~km~s$^{-1}$ are narrower than those detected in CO for W44 \cite[$\Delta v>25$~km~s$^{-1}$;][]{seta04,anderl14}, W28 \cite[$\Delta v = 20-30$~km~s$^{-1}$;][]{reach05}, and G 357.7+0.3 \cite[$\Delta v =15-30$~km~s$^{-1}$;][]{rho17}, and closer to some clumps in IC~443 \cite[$\Delta v \approx 15$~km~s$^{-1}$;][]{vandishoeck93} and HB~3 \cite[$\Delta v = 8-20$~km~s$^{-1}$;][]{rho21}; we note, of course, that these are line-of-sight projections. Whether these molecules are shock heated without being dissociated, or partially dissociated and re-formed behind the shock, is unclear.

In the case of G 350.0-2.0, as discussed, we find no evidence of interaction, in spite of the complete coverage of the SNR area, and the presence of an abundance of molecular gas. But absence of evidence is not evidence of absence; interaction signatures could be hidden due to line-of-sight complications such as the presence of absorption, or due to the relative faintness of the shocked material compared with the unshocked gas. We have not ruled out, either for VRO 42.05.01 nor for G 350.0-2.0, that a density gradient caused the wing structure. However, the truly special morphological feature of these two sources --the concavity of the two corners where the shell meets the wing-- seems to beg for a density discontinuity, rather than simply a gradient, that we just cannot see in dense molecular gas (but in the case of VRO 42.05.01 could be accounted for by the dust scarf mentioned in section \ref{sec:ir}). 

We believe that the morphology of these two shell and wing sources is different from that of sources showing a shock breakout into a lower density environment. Consider the case of CTB 1 \citep{pannuti10,katsuragawa18}: this source shows as a circular bubble in radio, and, most clearly, optical emission \cite[see fig. 9 in][]{pannuti10}, with a chimney-line breakout that the X-ray emission extends through, and well behind. \cite{pannuti10} interpret CTB 1 as the result of an explosion that later broke out into a low-density bubble, which certainly seems in line with radio, optical, and X-ray observations. This is rather different from the morphology of VRO 42.05.01 and G 350.0-2.0: the bubble is circular, whereas the wings are sharp triangles (this observation initially motivated our suggestion that the wing is due to an explosion into the cavity created by a bow shock), and the breakout into a chimney is open, as opposed to the confined dome of the shells.

\section{Conclusions}

We performed CO observations towards two SNRs that show a shell and wing shape: VRO 42.05.01 and G 350.0-2.0. We observed VRO 42.05.01 with the IRAM 30m telescope in the \twcooz, \twcoto, and \thcooz\ transitions, and located three regions with large velocity dispersions ($\Delta v > 10$~km~s$^{-1}$) and a \twcoto\ to \twcooz\ ratio larger than 1.8. These are signatures of heated and compressed gas, and we interpret the observations as evidence that the molecular gas has been shocked by the SNR forward shock. We also found several regions with swept up molecular gas (but without enhanced CO ratios), and several regions exhibiting evidence of an interaction with a less strong shock ($\Delta v < -4.5$~km~s$^{-1}$, but CO ratios larger than one), some of which, we propose, could be due to a stellar wind pushing and/or shock heating the molecular material. We observed SNR G 350.0-2.0 with the APEX telescope in the \twcoto\ transition, and the only hints of interaction that we found are asymmetrical wings in three of the line profiles, and some unresolved indication of arcs in position-velocity space. These constitute poor evidence of interaction, and more observations are required in this direction to confirm or reject association between the SNR and its molecular environment.

We further used archival data to compare these two sources. We used data from the GLEAM survey to understand the low radio frequency properties of G 350.0-2.0, and found that the spectral index remains flat as measured for higher frequencies up to 200~MHz ($\alpha\approx-0.4$) but steepens to $\alpha\approx-0.7$ between 88~MHz and 200~MHz. This is similar to what we had observed with our earlier LOFAR observations of VRO 42.05.01 \citep{arias19b}. On the other hand, the infrared and optical environments of these sources are rather different --whereas for VRO~42.05.01 there is a lot of correspondence between the optical, infrared, and radio emission, this is not the case for G 350.0-2.0 (although it bears noting that the latter has not been observed in [S~\textsc{ii}] and [O~\textsc{iii}]). 

The observed environmental differences between these two SNRs suggest that the shell and wing morphology they exhibit is not due to large-scale environmental effects, at least as pertains to their molecular environments (a large-scale density gradient that remains undetected in both cases is still a possibility). It is plausible, therefore, that the morphology could be due to common stellar evolution properties of their progenitors stars, such as the circumstellar environment carved by a mass-losing runaway progenitor. An interesting potential implication is that the mixed morphology of these sources (the internal X-ray emission surrounded by a radio shell), could also be due to interactions with the material the stars shed during their lifetimes. This is quite a different outlook from what traditional mixed morphology models have invoked. 

\begin{acknowledgements}
We thank the referee, J. Rho, for her thoughtful feedback on this paper.
This work is based on observations carried out under project number 045-18 159-19 with the IRAM 30m telescope. IRAM is supported by INSU/CNRS (France), MPG (Germany) and IGN (Spain). It is also based on data acquired with the Atacama Pathfinder Experiment (APEX) under programme ID 105.209F. APEX is a collaboration between the Max-Planck-Institut fur Radioastronomie, the European Southern Observatory,
and the Onsala Space Observatory. We thank the staff of IRAM and APEX for their support with the observations.

We thank B. Gaensler for sharing his radio FITS file of G 350.0-2.0.  M.A. acknowledges support from the VENI research programme with project number 202.143, which is financed by the Netherlands Organisation for Scientific Research (NWO). P.Z. acknowledges the support from NSFC grant No. 12273010 and Nederlandse Organisatie voor Wetenschappelijk Onderzoek (NWO) Veni Fellowship, grant no. 639.041.647. The research leading to these results has received funding from the European Union’s Horizon 2020 research and innovation program under RadioNet.

\end{acknowledgements}

\bibliographystyle{aa} 
\bibliography{biblio.bib} 

\begin{thebibliography}{56}
\expandafter\ifx\csname natexlab\endcsname\relax\def\natexlab#1{#1}\fi

\bibitem[{{Anderl} {et~al.}(2014){Anderl}, {Gusdorf}, \&
  {G{\"u}sten}}]{anderl14}
{Anderl}, S., {Gusdorf}, A., \& {G{\"u}sten}, R. 2014, \aap, 569, A81

\bibitem[{{Araya}(2013)}]{araya13}
{Araya}, M. 2013, \mnras, 434, 2202

\bibitem[{{Arias} {et~al.}(2019{\natexlab{a}}){Arias}, {Dom{\v{c}}ek}, {Zhou},
  \& {Vink}}]{arias19}
{Arias}, M., {Dom{\v{c}}ek}, V., {Zhou}, P., \& {Vink}, J. 2019{\natexlab{a}},
  \aap, 627, A75

\bibitem[{{Arias} {et~al.}(2019{\natexlab{b}}){Arias}, {Vink}, {Iacobelli},
  {Dom{\v{c}}ek}, {Haverkorn}, {Oonk}, {Polderman}, {Reich}, {White}, \&
  {Zhou}}]{arias19b}
{Arias}, M., {Vink}, J., {Iacobelli}, M., {et~al.} 2019{\natexlab{b}}, \aap,
  622, A6

\bibitem[{{Burrows} \& {Guo}(1994)}]{burrows94}
{Burrows}, D.~N. \& {Guo}, Z. 1994, \apjl, 421, L19

\bibitem[{{Bykov} {et~al.}(2008){Bykov}, {Krassilchtchikov}, {Uvarov},
  {Bloemen}, {Bocchino}, {Dubner}, {Giacani}, \& {Pavlov}}]{bykov08}
{Bykov}, A.~M., {Krassilchtchikov}, A.~M., {Uvarov}, Y.~A., {et~al.} 2008,
  \apj, 676, 1050

\bibitem[{{Chen} {et~al.}(2008){Chen}, {Seward}, {Sun}, \& {Li}}]{chen08}
{Chen}, Y., {Seward}, F.~D., {Sun}, M., \& {Li}, J.-t. 2008, \apj, 676, 1040

\bibitem[{{Chen} {et~al.}(2013){Chen}, {Zhou}, \& {Chu}}]{chen13}
{Chen}, Y., {Zhou}, P., \& {Chu}, Y.-H. 2013, \apjl, 769, L16

\bibitem[{{Chiotellis} {et~al.}(2019){Chiotellis}, {Boumis}, {Derlopa}, \&
  {Steffen}}]{Chiotellis2019}
{Chiotellis}, A., {Boumis}, P., {Derlopa}, S., \& {Steffen}, W. 2019, arXiv
  e-prints, arXiv:1909.08947

\bibitem[{{Choe} \& {Jung}(1997)}]{seung-urn97}
{Choe}, S.-U. \& {Jung}, H.-C. 1997, Publication of Korean Astronomical
  Society, 12, 173

\bibitem[{{Claussen} {et~al.}(1997){Claussen}, {Frail}, {Goss}, \&
  {Gaume}}]{claussen97}
{Claussen}, M.~J., {Frail}, D.~A., {Goss}, W.~M., \& {Gaume}, R.~A. 1997, \apj,
  489, 143

\bibitem[{{Cornett} {et~al.}(1977){Cornett}, {Chin}, \& {Knapp}}]{cornett77}
{Cornett}, R.~H., {Chin}, G., \& {Knapp}, G.~R. 1977, \aap, 54, 889

\bibitem[{{Cox} {et~al.}(1999){Cox}, {Shelton}, {Maciejewski}, {Smith},
  {Plewa}, {Pawl}, \& {R{\'o}{\.z}yczka}}]{cox99}
{Cox}, D.~P., {Shelton}, R.~L., {Maciejewski}, W., {et~al.} 1999, \apj, 524,
  179

\bibitem[{{Dalgarno} \& {McCray}(1972)}]{dalgarno72}
{Dalgarno}, A. \& {McCray}, R.~A. 1972, \araa, 10, 375

\bibitem[{{Denoyer}(1979)}]{denoyer79}
{Denoyer}, L.~K. 1979, \apjl, 232, L165

\bibitem[{{Derlopa} {et~al.}(2020){Derlopa}, {Boumis}, {Chiotellis}, {Steffen},
  \& {Akras}}]{derlopa20}
{Derlopa}, S., {Boumis}, P., {Chiotellis}, A., {Steffen}, W., \& {Akras}, S.
  2020, \mnras, 499, 5410

\bibitem[{{Fesen} {et~al.}(1983){Fesen}, {Gull}, \& {Ketelsen}}]{fesen83}
{Fesen}, R.~A., {Gull}, T.~R., \& {Ketelsen}, D.~A. 1983, \apjs, 51, 337

\bibitem[{{Gaensler}(1998)}]{gaensler98}
{Gaensler}, B.~M. 1998, \apj, 493, 781

\bibitem[{{Guo} \& {Burrows}(1997)}]{guo97}
{Guo}, Z. \& {Burrows}, D.~N. 1997, \apjl, 480, L51

\bibitem[{{Hurley-Walker} {et~al.}(2019){Hurley-Walker}, {Hancock}, {Franzen},
  {Callingham}, {Offringa}, {Hindson}, {Wu}, {Bell}, {For}, {Gaensler},
  {Johnston-Hollitt}, {Kapi{\'n}ska}, {Morgan}, {Murphy}, {McKinley},
  {Procopio}, {Staveley-Smith}, {Wayth}, \& {Zheng}}]{hurley-walker19}
{Hurley-Walker}, N., {Hancock}, P.~J., {Franzen}, T.~M.~O., {et~al.} 2019,
  \pasa, 36, e047

\bibitem[{{Jiang} {et~al.}(2010){Jiang}, {Chen}, {Wang}, {Su}, {Zhou},
  {Safi-Harb}, \& {DeLaney}}]{jiang10}
{Jiang}, B., {Chen}, Y., {Wang}, J., {et~al.} 2010, \apj, 712, 1147

\bibitem[{{Karpova} {et~al.}(2016){Karpova}, {Shternin}, {Zyuzin}, {Danilenko},
  \& {Shibanov}}]{karpova16}
{Karpova}, A., {Shternin}, P., {Zyuzin}, D., {Danilenko}, A., \& {Shibanov}, Y.
  2016, \mnras, 462, 3845

\bibitem[{{Katsuragawa} {et~al.}(2018){Katsuragawa}, {Nakashima}, {Matsumura},
  {Tanaka}, {Uchida}, {Lee}, {Uchiyama}, {Arakawa}, \&
  {Takahashi}}]{katsuragawa18}
{Katsuragawa}, M., {Nakashima}, S., {Matsumura}, H., {et~al.} 2018, \pasj, 70,
  110

\bibitem[{{Landecker} {et~al.}(1982){Landecker}, {Pineault}, {Routledge}, \&
  {Vaneldik}}]{landercker82}
{Landecker}, T.~L., {Pineault}, S., {Routledge}, D., \& {Vaneldik}, J.~F. 1982,
  \apjl, 261, L41

\bibitem[{{Landecker} {et~al.}(1989){Landecker}, {Pineault}, {Routledge}, \&
  {Vaneldik}}]{landecker89}
{Landecker}, T.~L., {Pineault}, S., {Routledge}, D., \& {Vaneldik}, J.~F. 1989,
  \mnras, 237, 277

\bibitem[{{Leahy} \& {Tian}(2005)}]{leahy05}
{Leahy}, D. \& {Tian}, W. 2005, \aap, 440, 929

\bibitem[{{Lozinskaia}(1979)}]{lozinskaia79}
{Lozinskaia}, T.~A. 1979, Australian Journal of Physics, 32, 113

\bibitem[{{Ma{\'\i}z Apell{\'a}niz} {et~al.}(2018){Ma{\'\i}z Apell{\'a}niz},
  {Pantaleoni Gonz{\'a}lez}, {Barb{\'a}}, {Sim{\'o}n-D{\'\i}az}, {Negueruela},
  {Lennon}, {Sota}, \& {Trigueros P{\'a}ez}}]{maizapellaniz18}
{Ma{\'\i}z Apell{\'a}niz}, J., {Pantaleoni Gonz{\'a}lez}, M., {Barb{\'a}},
  R.~H., {et~al.} 2018, \aap, 616, A149

\bibitem[{{Matsumura} {et~al.}(2017){Matsumura}, {Uchida}, {Tanaka}, {Tsuru},
  {Nobukawa}, {Nobukawa}, \& {Itou}}]{matsumura17}
{Matsumura}, H., {Uchida}, H., {Tanaka}, T., {et~al.} 2017, \pasj, 69, 30

\bibitem[{{Meyer} {et~al.}(2015){Meyer}, {Langer}, {Mackey}, {Vel{\'a}zquez},
  \& {Gusdorf}}]{meyer15}
{Meyer}, D.~M.~A., {Langer}, N., {Mackey}, J., {Vel{\'a}zquez}, P.~F., \&
  {Gusdorf}, A. 2015, \mnras, 450, 3080

\bibitem[{{Meyer} {et~al.}(2023){Meyer}, {Pohl}, {Petrov}, \&
  {Egberts}}]{meyer23}
{Meyer}, D.~M.~A., {Pohl}, M., {Petrov}, M., \& {Egberts}, K. 2023, \mnras,
  521, 5354

\bibitem[{{Ohnishi} {et~al.}(2011){Ohnishi}, {Koyama}, {Tsuru}, {Masai},
  {Yamaguchi}, \& {Ozawa}}]{ohnishi11}
{Ohnishi}, T., {Koyama}, K., {Tsuru}, T.~G., {et~al.} 2011, \pasj, 63, 527

\bibitem[{{Pannuti} {et~al.}(2010){Pannuti}, {Rho}, {Borkowski}, \&
  {Cameron}}]{pannuti10}
{Pannuti}, T.~G., {Rho}, J., {Borkowski}, K.~J., \& {Cameron}, P.~B. 2010, \aj,
  140, 1787

\bibitem[{{Pineault} {et~al.}(1987){Pineault}, {Landecker}, \&
  {Routledge}}]{pineault87}
{Pineault}, S., {Landecker}, T.~L., \& {Routledge}, D. 1987, \apj, 315, 580

\bibitem[{{Pineault} {et~al.}(1985){Pineault}, {Pritchet}, {Landecker},
  {Routledge}, \& {Vaneldik}}]{pineault85}
{Pineault}, S., {Pritchet}, C.~J., {Landecker}, T.~L., {Routledge}, D., \&
  {Vaneldik}, J.~F. 1985, \aap, 151, 52

\bibitem[{{Reach} \& {Rho}(1999)}]{reach99}
{Reach}, W.~T. \& {Rho}, J. 1999, \apj, 511, 836

\bibitem[{{Reach} {et~al.}(2005){Reach}, {Rho}, \& {Jarrett}}]{reach05}
{Reach}, W.~T., {Rho}, J., \& {Jarrett}, T.~H. 2005, \apj, 618, 297

\bibitem[{{Reid} {et~al.}(2014){Reid}, {Menten}, {Brunthaler}, {Zheng}, {Dame},
  {Xu}, {Wu}, {Zhang}, {Sanna}, {Sato}, {Hachisuka}, {Choi}, {Immer},
  {Moscadelli}, {Rygl}, \& {Bartkiewicz}}]{reid14}
{Reid}, M.~J., {Menten}, K.~M., {Brunthaler}, A., {et~al.} 2014, \apj, 783, 130

\bibitem[{{Rho} {et~al.}(2017){Rho}, {Hewitt}, {Bieging}, {Reach}, {Andersen},
  \& {G{\"u}sten}}]{rho17}
{Rho}, J., {Hewitt}, J.~W., {Bieging}, J., {et~al.} 2017, \apj, 834, 12

\bibitem[{{Rho} {et~al.}(2021){Rho}, {Jarrett}, {Tram}, {Lim}, {Reach},
  {Bieging}, {Lee}, {Koo}, \& {Whitney}}]{rho21}
{Rho}, J., {Jarrett}, T.~H., {Tram}, L.~N., {et~al.} 2021, \apj, 917, 47

\bibitem[{{Rho} \& {Petre}(1998)}]{rho98}
{Rho}, J. \& {Petre}, R. 1998, \apjl, 503, L167

\bibitem[{{Scaife} \& {Heald}(2012)}]{scaife12}
{Scaife}, A. M.~M. \& {Heald}, G.~H. 2012, \mnras, 423, L30

\bibitem[{{Seta} {et~al.}(1998){Seta}, {Hasegawa}, {Dame}, {Sakamoto}, {Oka},
  {Handa}, {Hayashi}, {Morino}, {Sorai}, \& {Usuda}}]{seta98}
{Seta}, M., {Hasegawa}, T., {Dame}, T.~M., {et~al.} 1998, \apj, 505, 286

\bibitem[{{Seta} {et~al.}(2004){Seta}, {Hasegawa}, {Sakamoto}, {Oka}, {Sawada},
  {Inutsuka}, {Koyama}, \& {Hayashi}}]{seta04}
{Seta}, M., {Hasegawa}, T., {Sakamoto}, S., {et~al.} 2004, \aj, 127, 1098

\bibitem[{{Shelton} {et~al.}(1999){Shelton}, {Cox}, {Maciejewski}, {Smith},
  {Plewa}, {Pawl}, \& {R{\'o}{\.z}yczka}}]{shelton99}
{Shelton}, R.~L., {Cox}, D.~P., {Maciejewski}, W., {et~al.} 1999, \apj, 524,
  192

\bibitem[{{Stupar} \& {Parker}(2011)}]{stupar11}
{Stupar}, M. \& {Parker}, Q.~A. 2011, \mnras, 414, 2282

\bibitem[{{Taylor} {et~al.}(2003){Taylor}, {Gibson}, {Peracaula}, {Martin},
  {Landecker}, {Brunt}, {Dewdney}, {Dougherty}, {Gray}, {Higgs}, {Kerton},
  {Knee}, {Kothes}, {Purton}, {Uyaniker}, {Wallace}, {Willis}, \&
  {Durand}}]{taylor03}
{Taylor}, A.~R., {Gibson}, S.~J., {Peracaula}, M., {et~al.} 2003, \aj, 125,
  3145

\bibitem[{{Treffers}(1979)}]{treffers79}
{Treffers}, R.~R. 1979, \apjl, 233, L17

\bibitem[{{Ustamujic} {et~al.}(2021){Ustamujic}, {Orlando}, {Greco}, {Miceli},
  {Bocchino}, {Tutone}, \& {Peres}}]{ustamujic21}
{Ustamujic}, S., {Orlando}, S., {Greco}, E., {et~al.} 2021, \aap, 649, A14

\bibitem[{{van der Tak} {et~al.}(2007){van der Tak}, {Black}, {Sch{\"o}ier},
  {Jansen}, \& {van Dishoeck}}]{radex}
{van der Tak}, F.~F.~S., {Black}, J.~H., {Sch{\"o}ier}, F.~L., {Jansen}, D.~J.,
  \& {van Dishoeck}, E.~F. 2007, \aap, 468, 627

\bibitem[{{van Dishoeck} {et~al.}(1993){van Dishoeck}, {Jansen}, \&
  {Phillips}}]{vandishoeck93}
{van Dishoeck}, E.~F., {Jansen}, D.~J., \& {Phillips}, T.~G. 1993, \aap, 279,
  541

\bibitem[{{Vink}(2012)}]{vink12}
{Vink}, J. 2012, \aapr, 20, 49

\bibitem[{{Voges} {et~al.}(1999){Voges}, {Aschenbach}, {Boller},
  {Br{\"a}uninger}, {Briel}, {Burkert}, {Dennerl}, {Englhauser}, {Gruber},
  {Haberl}, {Hartner}, {Hasinger}, {K{\"u}rster}, {Pfeffermann}, {Pietsch},
  {Predehl}, {Rosso}, {Schmitt}, {Tr{\"u}mper}, \& {Zimmermann}}]{voges99}
{Voges}, W., {Aschenbach}, B., {Boller}, T., {et~al.} 1999, \aap, 349, 389

\bibitem[{{White} \& {Long}(1991)}]{white91}
{White}, R.~L. \& {Long}, K.~S. 1991, \apj, 373, 543

\bibitem[{{Wright} {et~al.}(2010){Wright}, {Eisenhardt}, {Mainzer}, {Ressler},
  {Cutri}, {Jarrett}, {Kirkpatrick}, {Padgett}, {McMillan}, {Skrutskie},
  {Stanford}, {Cohen}, {Walker}, {Mather}, {Leisawitz}, {Gautier}, {McLean},
  {Benford}, {Lonsdale}, {Blain}, {Mendez}, {Irace}, {Duval}, {Liu}, {Royer},
  {Heinrichsen}, {Howard}, {Shannon}, {Kendall}, {Walsh}, {Larsen}, {Cardon},
  {Schick}, {Schwalm}, {Abid}, {Fabinsky}, {Naes}, \& {Tsai}}]{wright10}
{Wright}, E.~L., {Eisenhardt}, P. R.~M., {Mainzer}, A.~K., {et~al.} 2010, \aj,
  140, 1868

\bibitem[{{Xiao} {et~al.}(2022){Xiao}, {Zhu}, {Sun}, {Jiang}, \&
  {Sun}}]{xiao22}
{Xiao}, L., {Zhu}, M., {Sun}, X.-H., {Jiang}, P., \& {Sun}, C. 2022, Research
  in Astronomy and Astrophysics, 22, 035003

\end{thebibliography}
%

\end{document}